\documentclass[fleqn,usenatbib]{mnras}

\usepackage{newtxtext,newtxmath}

\usepackage[T1]{fontenc}

\DeclareRobustCommand{\VAN}[3]{#2}
\let\VANthebibliography\thebibliography
\def\thebibliography{\DeclareRobustCommand{\VAN}[3]{##3}\VANthebibliography}

\usepackage{graphicx}	
\usepackage{amsmath, bm}	

\usepackage{cellspace}
\usepackage{array}
\newcolumntype{P}[1]{>{\centering\arraybackslash}p{#1}}
\newcolumntype{D}{ >{\centering\arraybackslash} c{1cm} }
\usepackage{transparent}
\usepackage{comment}
\newcommand{\Hquad}{\hspace{0.5em}}
\usepackage[table, dvipsnames]{xcolor}
\usepackage{float}
\usepackage{subcaption}
\definecolor{olivine}{rgb}{0.6, 0.73, 0.45}

\title[$M_{\mathrm{d}} - M_{\star}$ and $\dot M - M_{\star}$ correlations in protoplanetary discs]{On the time evolution of the $M_{\mathrm{d}} - M_{\star}$ and $\dot M - M_{\star}$ correlations for protoplanetary discs: the viscous timescale increases with stellar mass}

\author[A. Somigliana et al.]{
Alice Somigliana$^{1, 2, 3}$\thanks{E-mail: alice.somigliana@eso.org},
Claudia Toci$^{2,4}$,
Giovanni Rosotti$^{5}$,
Giuseppe Lodato$^{2,4}$,
Marco Tazzari$^{6}$,
\newauthor
Carlo F. Manara$^{1}$,
Leonardo Testi$^{1,7}$ 
and Federico Lepri$^{2}$
\\
$^{1}$European Southern Observatory, Karl-Schwarzschild-Strasse 2, D-85748 Garching bei München, Germany \\
$^{2}$Dipartimento di Fisica, Università degli Studi di Milano, Via Celoria 16, 20133 Milano, Italy\\
$^{3}$Fakultat für Physik, Ludwig-Maximilians-Universität Munchen, Scheinerstr. 1, 81679 München, Germany\\
$^{4}$INAF - Osservatorio Astronomico di Brera, Via Brera 28, I-20121 Milano, Italy\\
$^{5}$School of Physics and Astronomy, University of Leicester, Leicester, LE1 7RH, United Kingdom\\
$^{6}$Institute of Astronomy, University of Cambridge, Madingley Road, CB3 0HA, Cambridge, UK\\
$^{7}$INAF-Osservatorio Astrofisico di Arcetri, Largo E. Fermi 5, I-50125 Firenze, Italy
}

\date{Accepted XXX. Received YYY; in original form ZZZ}

\pubyear{2022}

\begin{document}

\defcitealias{Ansdell+2017}{A17}
\defcitealias{Testi+2022-Ofiucone}{T22}
\defcitealias{Manara+2012}{M12}

\label{firstpage}
\pagerange{\pageref{firstpage}--\pageref{lastpage}}
\maketitle

\begin{abstract}
Large surveys of star-forming regions have
unveiled power-law correlations between the stellar mass and the disc parameters, such as the disc mass $M_{\mathrm{d}} \propto {M_{\star}}^{\lambda_{\mathrm{m}}}$ and the accretion rate $\dot M \propto {M_{\star}}^{\lambda_{\mathrm{acc}}}$. The observed slopes appear to be increasing with time, but the reason behind the establishment of these correlations and their subsequent evolution is still uncertain. We conduct a theoretical analysis of the impact of viscous evolution on power-law initial conditions for a population of protoplanetary discs. We find that, for evolved populations, viscous evolution enforces the two correlations to have the same slope, $\lambda_{\mathrm{m}}$ = $\lambda_{\mathrm{acc}}$, and that this limit is uniquely determined by the initial slopes $\lambda_{\mathrm{m}, 0}$ and $\lambda_{\mathrm{acc}, 0}$. We recover the increasing trend claimed from the observations when the difference in the initial values, $\delta_0 = \lambda_{\mathrm{m}, 0} - \lambda_{\mathrm{acc}, 0}$, is larger than $1/2$; moreover, we find that this increasing trend is a consequence of a positive correlation between the viscous timescale and the stellar mass. We also present the results of disc population synthesis numerical simulations, that allow us to introduce a spread and analyse the effect of sampling, which show a good agreement with our analytical predictions. Finally, we perform a preliminary comparison of our numerical results with observational data, which allows us to constrain the parameter space of the initial conditions to $\lambda_{\mathrm{m}, 0} \in [1.2, 2.1]$, $\lambda_{\mathrm{acc}, 0} \in [0.7, 1.5]$.
\end{abstract}




\section{Introduction} \label{sec:introduction}

Protoplanetary discs are the cradle of planets. Their evolution and dispersal strongly impact the outcome of planet formation; they especially affect the extent and availability of planetesimals, the building blocks of planets \citep{Morbidelli2012-planetformreview, Mordasini+2015}.

Discs also serve as a mass reservoir for the central accreting protostar. For accretion to take place, material stored in the disc needs to lose most of its angular momentum. The trigger to this process is conventionally identified as a macroscopic viscosity; the pioneering work of \cite{LyndenBellPringle1974}, based on the $\alpha$ prescription of \cite{ShakuraSunyaev1973}, set the ground for numerous following studies treating accretion as a redistribution of angular momentum within the disc. Despite being by far the most widely used, viscosity is not the only accretion theory; several studies have suggested MHD winds as promising candidates to explain protoplanetary disc accretion \citep{Lesur+2014, Bai2017simMHD, Bethune+2017, Lesur2020reviewMHD, Tabone+2021a, Tabone+2021b}. In this scenario, angular momentum is \textit{removed} instead of being redistributed, leading to significant differences in the evolutionary predictions.

Thanks to the great technological development of the last decades, and in particular to the advent of facilities like the Atacama Large Millimeter Array (ALMA), observational data allow to test evolutionary models. Extended data sets collecting information on a large number of young stellar objects provide the ideal ground to test theoretical predictions; the observational focus is therefore on surveys of entire star-forming regions. A number of such surveys have been already carried out (for example, \citealt{Barenfeld2016-masssurvey}, \citealt{Pascucci+2016}, \citealt{Ansdell+2016Survey}, \citealt{Testi+2016}, \citealt{Alcala+2017}, \citealt{Manara2017-Cha}, \citealt{Ansdell+2017}, \citealt{Cieza+2019-OphSurvey}, \citealt{Williams+2019-OphSurvey}, \citealt{Sanchis+2020Lupus}, \citealt{Testi+2022-Ofiucone}, \citealt{Manara+2020-UpperSco}), unveiling interesting features and patterns, such as power-law correlations between the properties of discs and their host stars.

Two crucial steps are required to test evolutionary models: first, performing numerical simulations of the different prescriptions; second, a comparison of these numerical results with data. Identifying key predictions for each model allows to distinguish between different scenarios. The viscous case shows a characteristic behaviour, known as viscous spreading: as part of the disc mass loses angular momentum and drifts inwards, accreting the protostar, another portion of the disc gains the same amount of angular momentum instead, expanding towards larger radii. Therefore, the radial extent of the disc increases with time, despite the ongoing accretion. This prediction does not apply to the MHD winds scenario \citep{Trapman+2021-MHDradii}; its removal of angular momentum causes disc radii to decrease as evolution proceeds. Whether the available data in the Lupus star-forming region agree with the viscous spreading predictions has been debated in the literature (\citealt{Sanchis+2020Lupus}, \citealt{Toci+2021}, \citealt{Trapman+2020-COradii}). Unfortunately, detecting viscous spreading using dust radius as a tracer is non-trivial: \cite{Rosotti+Letter2019} showed that this method requires significantly deep observations, targeting a fraction as high as 95$\%$ of the total flux.

The disc mass - accretion rate correlation provides another diagnostic criterion. In the purely viscous case, where the self-similar solution \citep{LyndenBellPringle1974} holds, such a correlation naturally stems from the analytical prescriptions for $M_\mathrm{d}$ and $\dot M$; in particular, it is expected to have slope of $\sim 1$ \citep{Hartmann1998, Mulders+2017, Lodato2017, Rosotti+2017} and a spread decreasing in time \citep{Lodato2017}, as the age of the population reaches and then outgrows the viscous timescale. The analysis of the first data from Lupus and Chameleon \citep{Manara2016-evidenceacc} showed a possible agreement with this prediction. However, \cite{Tabone+2021b} showed that an MHD disc winds model could be tuned to reproduce the $M_\mathrm{d} - \dot M$ correlation equally well, both in slope and spread - making it more challenging to distinguish between these two models solely using this relation. Moreover, there are some behaviours that cannot be fully explained by any of the two scenarios alone: an example is the Upper Scorpius star-forming region \citep{Manara+2020-UpperSco}, where the observational data show a large scatter in the $M_{\mathrm{d}} - \dot M$ relation (see also \citealt{Testi+2022-Ofiucone}). Additional mechanisms have been invoked to explain these inconsistencies, such as internal and external photoevaporation \citep{Rosotti+2017, Sellek+2020-extphoto, Somigliana2020} and dust evolution \citep{Sellek+2020-dustyorigin}; however, understanding the right combination of processes to retrieve the observed spread is non-trivial.

These similar behaviours, and the difficulties in observing the key prediction of viscous spreading, may sound discouraging. However, surveys also provide additional information - namely, the correlation between stellar and disc parameters. Many independent works \citep{Muzerolle+2003, Mohanty+2005, Natta+2006, Herczeg&Hillenbrand2008, Alcala+2014, Kalari+2015, Manara2017-Cha} have found a correlation between the disc accretion rate $\dot M$ and the stellar mass $M_{\star}$, with a slope of $\sim 1.8 \pm 0.2$; on the other hand, the disc mass versus stellar mass correlation appears to steepen with time (\citealt{Ansdell+2017}). Some attempts to explain both the $\dot M - M_{\star}$ correlation \citep{AlexanderArmitage2006, DullemondNattaTesti2006, Clarke+2006-MdotMstarinPMS, Ercolano+2014-MdotMstar} and the $M_\mathrm{d} - M_{\star}$ correlation \citep{Pascucci+2016, Pinilla+2020-MdiscMstar} have been made; nonetheless, it is sill not clear whether these correlations are determined by the initial conditions, or rather established later on in the disc lifetime as a consequence of the evolutionary processes. Whether the time evolution of these correlations can be understood in the context of viscously evolving disc populations has not been investigated so far.

The numerical counterpart of star-forming regions surveys is population syntheses, i.e., generating and evolving synthetic populations of discs through numerical methods. Performing population syntheses is particularly useful to test evolutionary models, as well as the impact of different physical effects on the diagnostic quantities of interest (such as disc masses and radii). Population syntheses have been carried out already, investigating different aspects of evolution and dispersal of protoplanetary discs \citep{Lodato2017, Somigliana2020, Sellek+2020-dustyorigin, Sellek+2020-extphoto}; however, none of them included both a proper Monte Carlo drawing of the involved parameters and the correlations between disc and stellar properties. The usual assumptions were a fixed stellar mass and a linear span of the parameter space, which make a good first approximation but lack the spread and statistics that really make population syntheses a powerful tool. In this paper, we employ a new and soon to be released Python code, \texttt{Diskpop}, which performs a coherent population synthesis of protoplanetary discs and sets the basis for future developments taking into account more and more physical effects acting on discs.

In this work, we aim to study the dependency of disc properties on the stellar mass $M_{\star}$, discussing its implications from the evolutionary point of view. These observed correlations are most likely linked to both evolution and initial conditions, and our goal is to disentangle between the two. We investigate the case where the correlations are already present as initial conditions. With this first paper we focus on setting up the framework for population syntheses: we limit our case study to purely viscous discs, but the natural progression is to include more evolutionary predictions to compare (and hopefully distinguish) between each other. The structure of the paper is as follows: in Section \ref{sec:summary_obs} we report and discuss state of the art of the observational evidences on disc masses, accretion rates and radii; in Section \ref{sec:analytical_considerations} we present our analytical considerations on the effects of viscous evolution on power-law initial correlations between the disc parameters and the stellar mass; in Section \ref{sec:population_synthesis} we discuss our population synthesis model, including both the implementation in \texttt{Diskpop} and the numerical results; we compare out results with observational data and discuss their implications in Section \ref{sec:discussion} and finally, we draw the conclusions of this work in Section \ref{sec:conclusions}.

\section{Summary of observational evidence} \label{sec:summary_obs}

\subsection{Disc mass} \label{subsec:disc_mass}

Numerous surveys of star-forming regions, where disc masses were determined observing the sub-mm continuum emission of the dust component \citep{Ansdell+2016Survey, Barenfeld2016-masssurvey, Pascucci+2016, Testi+2016, Ansdell+2017, Sanchis+2020Lupus, Testi+2022-Ofiucone}, have highlighted a power-law correlation between the disc mass $M_\mathrm{dust}$ (of the \textit{dusty} component) and the stellar mass $M_{\star}$. This relationship can be parametrised as linear in the logarithmic plane,

\begin{equation}
    \log_{10} \left( \frac{M_{\mathrm{dust}}}{M_{\oplus}} \right) = \lambda_{\mathrm{m}, \mathrm{obs}} \log_{10} \left( \frac{M_*}{M_{\odot}} \right) + q_{\mathrm{obs}} +  \varepsilon_{\mathrm{obs}},
    \label{eq:log_relation_mdiscmstar}
\end{equation}

\noindent with slope $\lambda_{\mathrm{m}, \mathrm{obs}}$ and intercept $q_{\mathrm{obs}}$\footnote{Our notation slightly differs from the ones previously used by \cite{Pascucci+2016} and \cite{Ansdell+2017}: in the first paper the slope (here $\lambda_{\mathrm{m}, \mathrm{obs}}$) is $\alpha$ and the intercept (here $q_{\mathrm{obs}}$) is $\beta$, while the second one uses the opposite convention. Moreover, we named the intrinsic dispersion $\Delta_{\mathrm{obs}}$ instead of $\delta$ to avoid confusion with another parameter that we define later in our paper.}. $\varepsilon_{\mathrm{obs}}$ is a gaussian random variable with mean 0, representing the scatter of the correlation.

The values of the parameters $\lambda_{\mathrm{m}, \mathrm{obs}}$ and $q_{\mathrm{obs}}$ can be determined from observations by fitting the correlation between $M_\mathrm{dust}$ and $M_{\star}$. \citet{Ansdell+2017} and \citet{Testi+2022-Ofiucone} (hereafter \citetalias{Ansdell+2017} and \citetalias{Testi+2022-Ofiucone} respectively) performed this fit for different star-forming regions; the results are shown in Table \ref{tab:Ansdell&Testi} (same as Table 4 in \citetalias{Ansdell+2017} and Table H.1 in \citetalias{Testi+2022-Ofiucone}), where $\Delta_{\mathrm{obs}}$ represents the intrinsic dispersion (namely, the standard deviation of the distribution of $\varepsilon_{\mathrm{obs}}$).

\begin{table}
\centering
\begin{tabular}{c c c c c}
    \hline
    \hline
    Region & Age [Myr] & $\lambda_{\mathrm{m}, \mathrm{obs}}$ & $q_{\mathrm{obs}}$ & $\Delta_{\mathrm{obs}}$\\
    \hline
    Taurus & 1-2 & 1.7 $\pm$ 0.2 & 1.2 $\pm$ 0.1 & 0.7 $\pm$ 0.1 \\
    Lupus & 1-3 & 1.8 $\pm$ 0.4 & 1.2 $\pm$ 0.2 & 0.9 $\pm$ 0.1 \\
    Cha I & 2-3 & 1.8 $\pm$ 0.3 & 1.0 $\pm$ 0.1 & 0.8 $\pm$ 0.1 \\
    $\sigma$ Orionis & 3-5 & 2.0 $\pm$ 0.4 & 1.0 $\pm$ 0.2 & 0.6 $\pm$ 0.1 \\
    Upper Sco & 5-11 & 2.4 $\pm$ 0.4 & 0.8 $\pm$ 0.2 & 0.7 $\pm$ 0.1 \\
    \hline
\end{tabular}

\vspace{0.5cm}

\begin{tabular}{c c c c c}
    \hline
    \hline
    Region & Median age [Myr] & $\lambda_{\mathrm{m}, \mathrm{obs}}$ & $q_{\mathrm{obs}}$ & $\Delta_{\mathrm{obs}}$\\
    \hline
    Corona & 0.6 & 1.3 $\pm$ 0.5 & 0.4 $\pm$ 0.4 & 1.1 $\pm$ 0.7 \\
    Taurus & 0.9 & 1.5 $\pm$ 0.2 & 1.1 $\pm$ 0.1 & 0.8 $\pm$ 0.3 \\
    L1668 & 1 & 1.5 $\pm$ 0.2 & 1.0 $\pm$ 0.1 & 0.8 $\pm$ 0.3 \\
    Lupus & 2 & 1.7 $\pm$ 0.3 & 1.4 $\pm$ 0.2 & 0.7 $\pm$ 0.3 \\
    Cha I & 2.8 & 1.6 $\pm$ 0.3 & 1.1 $\pm$ 0.2 & 0.7 $\pm$ 0.4 \\
    Upper Sco & 4.3 & 2.2 $\pm$ 0.3 & 0.8 $\pm$ 0.2 & 0.7 $\pm$ 0.3 \\
    \hline
\end{tabular}
\caption{Fitted values of $\lambda_{\mathrm{m}, \mathrm{obs}}$, $q_{\mathrm{obs}}$ and $\Delta_{\mathrm{obs}}$ (see paragraph \ref{subsec:disc_mass} for details) for different star-forming regions as performed by \citetalias{Ansdell+2017} (top table) and \protect\citetalias{Testi+2022-Ofiucone} (bottom table).}
\label{tab:Ansdell&Testi}
\end{table}

Table \ref{tab:Ansdell&Testi} shows that the mean value of $q_{\mathrm{obs}}$ tends to get lower and lower with time. This behaviour, which is more visible in the top panel, is an intrinsic characteristic of the standard viscous scenario: as discs get older, part of their mass is lost due to the ongoing accretion of the central protostar, which eventually depletes the disc. However, it is worth pointing out that some star-forming regions appear not to follow this trend \citep{Cazzoletti2019, Williams+2019-OphSurvey}, and the reason behind that is still unclear.

On the other hand, the mean value of $\lambda_{\mathrm{m}, \mathrm{obs}}$ appears to be increasing with time, implying a steepening of the correlation between the (dust) disc mass and the stellar mass. Investigating the mathematical origin, physical meaning and expected evolution of such a trend is one of the main goals of this paper, and will be addressed from Section \ref{sec:analytical_considerations}.

\subsection{Disc accretion rate} \label{subsec:disc_accrate}

The main signatures of young stars accreting material from their surrounding disc can be found in their spectra. Gas falling onto the stellar surface along the magnetic field lines \citep{CalvetGullbring1998} causes an excess emission, paticularly visible in the UV area of the spectrum (and especially in the Balmer continuum, see \citealt{Gullbring1998}). Characteristic emission line profiles are also typical indicators of accretion. Modelling the Balmer continuum excess in the spectra of young stars, and fitting emission line profiles, provide effective ways of measuring accretion rates.

Numerous surveys focusing on different star-forming regions have targeted accretion rates \citep{Muzerolle+2003, Natta+2004-lowmassaccrates, Mohanty+2005, DullemondNattaTesti2006, Herczeg&Hillenbrand2008, Rigliaco+2011, Manara+2012, Alcala+2014, Alcala+2017, Kalari+2015, Manara+2016-ChaI, Manara2017-Cha, Manara+2020-UpperSco, Testi+2022-Ofiucone}. Many of them have found a power-law correlation between the accretion rate and the stellar mass, $\dot M \propto {M_{\star}}^{\lambda_{\mathrm{acc}, \mathrm{obs}}}$. The best fit value of $\lambda_{\mathrm{acc}, \mathrm{obs}} \approx 1.8 \pm 0.2$ seems to be roughly constant throughout different regions, suggesting that it could be independent on age (unlike the $M_\mathrm{d} - M_{\star}$ correlation, see subsection \ref{subsec:disc_mass}). On the contrary, \citet{Manara+2012} (hereafter \citetalias{Manara+2012}) do see an increasing trend of $\lambda_{\mathrm{acc}, \mathrm{obs}}$ with the age of the population, a trend similar to the one claimed by \citetalias{Ansdell+2017} with respect to $\lambda_{\mathrm{m}, \mathrm{obs}}$. There is, however, a significant difference between this latter work and the others mentioned: \citetalias{Manara+2012} analysed a sample of $\sim 700$ stars in the single star-forming region of the Orion Nebula Cluster, determining the isochronal age of each object in the region independently. On the other hand, \citetalias{Ansdell+2017} and \citetalias{Testi+2022-Ofiucone} considered different star-forming regions and assumed all of the objects in each of them to be coeval - determining therefore the mean age of each sample. The ages of young stellar objects are usually determined by comparing their position on the Hertzsprung-Russell (H-R) diagram with theoretical isochrones (see \citealt{Soderblom+2014PPVI} for a review): however, this method comes with a series of caveats \citep{Preibisch2021ReviewAges}. In particular, translating a spread in luminosity into a spread in ages is not straightforward due to a number of factors that can impact the shape of the H-R diagram, such as measurement uncertainties and variation of the accretion processes. Determining a mean age for a whole star-forming region absorbs part of this uncertainty, which is the reason why the approach of \citetalias{Manara+2012} is less used. Nonetheless, their results intriguingly show a behaviour of $\lambda_{\mathrm{acc}, \mathrm{obs}}$ similar to that of $\lambda_{\mathrm{m}, \mathrm{obs}}$, and therefore represent a case study worth considering. In this work, when comparing to observational data (see subsection \ref{subsec:obsdata}), we will consider the fits for $\lambda_{\mathrm{acc}, \mathrm{obs}}$ obtained from \citetalias{Manara+2012} as well as \citetalias{Testi+2022-Ofiucone}. Table \ref{tab:lambda2_fits} summarises the fitted values from both works.

\begin{table}
\begin{minipage}{0.47\linewidth}
    \centering
    \begin{tabular}{c c}
         \hline
         \hline
         Age [Myr] & $\lambda_{\mathrm{acc}, \mathrm{obs}}$ \\
         \hline
         $0.8$ & $1.15 \pm 2.00$ \\
         $1$ & $1.26 \pm 2.02$ \\
         $2$ & $1.61 \pm 2.06$ \\
         $5$ & $2.08 \pm 2.13$ \\
         $8$ & $2.32 \pm 2.16$ \\
         $10$ & $2.43 \pm 2.18$ \\
         \hline
    \end{tabular}
\end{minipage}
\hfill
\begin{minipage}{0.47\linewidth}
    \centering
    \begin{tabular}{c c c}
         \hline
         \hline
         Region & Age [Myr] & $\lambda_{\mathrm{acc}, \mathrm{obs}}$ \\
         \hline
         L1668 & $1.0$ & $1.8 \pm 0.5$ \\
         Lupus & $2.0$ & $1.6 \pm 0.3$ \\
         Cha I & $2.8$ & $2.3 \pm 0.3$ \\
         Upper Sco & $4.3$ & $1.5 \pm 0.8$ \\
         \hline
    \end{tabular}
\end{minipage}
    \caption{Fitted values of the $\dot M - M_{\star}$ slope, $\lambda_{\mathrm{acc}, \mathrm{obs}}$ (see paragraph \ref{subsec:disc_accrate} for details), by \protect\citetalias{Manara+2012} (left table) and \protect\citetalias{Testi+2022-Ofiucone} (right table)}
    \label{tab:lambda2_fits}
\end{table}

As discussed for $\lambda_{\mathrm{m}, \mathrm{obs}}$, there is in principle no theoretical reason for the correlation of the accretion rate with the stellar mass to steepen or flatten in time. However, if we assume the viscous framework to hold, we do expect the accretion rate to be a decreasing function of time; in particular, \cite{Hartmann1998} showed that the viscous evolution implies $\dot M \propto t^{- \eta}$, where $\eta \sim 1.5$. 

\subsection{Disc radius} \label{subsec:disc_radius}

Evolutionary models predict disc sizes to vary with time. In particular, radial drift is expected to influence the dust discs size \citep{Weidenschilling1977}: large dust grains drift inwards, eventually disappearing, while small grains are left behind and follow the motion of the gaseous component. In the viscous framework, gas is subject to the so-called viscous spreading: as a consequence of the conservation of angular momentum, the accretion onto the central star leads to a radial expansion of the disc. Once the large grains are removed, discs are expected to be wide and faint \citep{Rosotti+2019} and may be challenging to observe. These caveats must be taken into account when discussing disc radii, and in practice may severely limit the ability of observing viscous spreading, even when it does take place \citep{Toci+2021}.

The radial size of the dust component in discs is measured by analysing the extent of the millimetric thermal continuum emission. In the Ophiucus \citep{Cox+2017, Cieza+2019-OphSurvey}, Lupus \citep{Ansdell+2016Survey, Tazzari+2017, Andrews2018Sizes, Hendler2020} and Taurus \citep{Long+2019, Kurtovic+2021} star-forming regions, this method has been widely employed. \cite{Andrews2018Sizes} found evidence of a correlation between the dust disc radius and the stellar mass, namely $R_{\mathrm{d}} \propto {M_{\star}}^{0.6}$; recent works \citep{Andrews2018Sizes, Sanchis2021} have also found a correlation between the disc radius and the disc dust mass, which could be used to derive additional correlations with the stellar mass. However, as measuring radii requires to spatially resolve the discs, the sample of objects with measured $R_{\mathrm{d}}$ is smaller than that with measured $M_{\mathrm{d}}$; moreover, both those measurements carry significant uncertainties due to optically thick emission. This makes it not convenient to prefer this relation to $R_{\mathrm{d}} - M_{\star}$.

Using gas tracers, such as the rotational line emission of the $^{12}$CO molecule, one can also measure the gas disc size (R$_{\mathrm{CO}}$). As these observations are very time consuming, less data are available for the gaseous component \citep{Barenfeld2016-masssurvey, Ansdell+2018-radiigas, Sanchis2021}. In their work, \cite{Ansdell+2018-radiigas} did not see any correlation between the gas disc size and the stellar masses in Lupus, but their sample is biased towards the highest-masses discs around the highest-masses stars. At the present time, there is no strong evidence of a correlation between the disc gas size and the stellar mass; however, if a correlation exists, it is probably positive (see Appendix \ref{appendix-rdmstar}).

\section{Analytical considerations} \label{sec:analytical_considerations}

In the previous subsection, we have analysed the observational evidence for correlations between the the stellar mass and the three major disc properties - mass, accretion rate, and radius. We have presented the possibility, discussed by previous works, to describe these correlations (with the possible exception of $R_{\mathrm{d}} - M_{\star}$) with power-laws. In this section, we conduct a theoretical analysis of these correlations and their evolution in the purely viscous framework: we aim at determining the initial conditions needed to recover the observational findings.

We start with two assumptions:

\begin{enumerate}
    \item Self-similar discs: we assume that all discs in the population only evolve under the effect of the viscosity $\nu$. $\nu$ can be modelled as a power-law of the disc radius, $\nu = \nu_c \left( \frac{R}{R_c} \right)^{\gamma}$, where $\nu_c = \nu (R=R_c)$ and $R_c$ is the exponential cutoff radius. The value of $\nu$ is prescribed following \cite{ShakuraSunyaev1973} as $\nu = \alpha c_s H$ (where $\alpha$ is a dimensionless parameter, $c_s$ is the sound speed, $H$ the height of the disc). This implies that the self-similar solution by \cite{LyndenBellPringle1974} describes the evolution of the gas component in protoplanetary discs,
    
    \begin{equation}
        \Sigma(R, t) = \frac{M_{\mathrm{d}}(0)}{2 \pi {R_c}^2}(2-\gamma) \left( \frac{R}{R_c} \right)^{-\gamma} T^{-\eta} \exp \left( - \frac{(R/R_c)^{2-\gamma}}{T} \right),
        \label{eq:selfsimilarsol}
    \end{equation}
    
    \noindent where $\eta = (5/2 - \gamma)/(2 - \gamma)$, $T = 1 + t/t_{\nu}$ and $t_{\nu}$ is the viscous timescale, defined as $t_{\nu} = {R_c}^2/(3 \nu_c)$, on which viscous processes leading to evolution of discs take place.

    \item Power-law initial conditions: we assume power-law correlations with the stellar mass as \textit{initial conditions} for $M_{\mathrm{d}}$, $\dot M$ and $R_c$. The value of the slopes can vary as evolution takes place, and we will refer to the evolved values as $\lambda_{\mathrm{m}}$, $\lambda_{\mathrm{acc}}$ and $\zeta$. The initial correlations are set as follows:
    
    \begin{equation}
        \begin{cases}
            M_\mathrm{d} (0) \propto {M_{\star}}^{\lambda_{\mathrm{m}, 0}} \\
            \dot{M} (0) \propto {M_{\star}}^{\lambda_{\mathrm{acc}, 0}} \\
            R_c (0) \propto {M_{\star}}^{\zeta_0}.
        \end{cases}
    \label{eq:assumptions}
    \end{equation}
\end{enumerate}

The self-similar solution provides analytical expressions for both the disc mass and accretion rate:

\begin{equation}
    M_\mathrm{d} (t) = M_{\mathrm{d}, 0} \left( 1 + \frac{t}{t_{\nu}} \right)^{1 - \eta},
    \label{eq:discmass_ss}
\end{equation}

\begin{equation}
    \dot M (t) = (\eta - 1) \frac{M_{\mathrm{d}, 0}}{t_{\nu}} \left( 1 + \frac{t}{t_{\nu}} \right)^{- \eta}.
    \label{eq:accrate_ss}
\end{equation}

\noindent Note that, as the ratio $M_{\mathrm{d}} / \dot M$ has the dimension of a time, by choosing the initial $M_{\mathrm{d}}$ and $\dot M$ we completely specify their time evolution. Depending on the age of the disc with respect to its viscous timescale, set by the radius of the disc itself, discs can be considered young ($t \ll t_{\nu}$) or evolved ($t \gg t_{\nu}$). In these two limits, it is possible to simplify equations \eqref{eq:discmass_ss} and \eqref{eq:accrate_ss}. We divide the following discussion based on these two evolutionary scenarios, as they lead to different results and theoretical expectations (subsections \ref{subsec:young_pop} and \ref{subsec:evolved_pop}).

Equations \eqref{eq:discmass_ss} and \eqref{eq:accrate_ss} make clear that the time evolution of the slopes depends on how $t_{\nu}$ scales with stellar mass. Because at $t = 0$ the initial accretion rate can be written as $\dot M_0 \propto M_{\mathrm{d}, 0}/t_{\nu}$, a scaling of $t_{\nu}$ is already implicitly assumed given a choice of $\lambda_{\mathrm{m},0}$ and $\lambda_{\mathrm{acc},0}$. We can see from the definition of the viscous timescale that the dependence on stellar mass is contained in any of the three parameters $\alpha$, $c_s$, or $H$. In this paper, we assume $\alpha$ to be a constant across all stellar masses; moreover, we also consider it as constant in radius and time, and fix its value to $10^{-3}$. However, note that the assumption that $\alpha$ does not depend on stellar mass does not affect our general results on the evolution of the slopes (see the end of subsection \ref{subsec:evolved_pop}), since this depends only on the scaling of viscous timescale with stellar mass. On the other hand, this assumption does affect how aspect ratio $H/R$ and disc radius scale with stellar mass. Regarding the aspect ratio, this quantity can be defined as

\begin{equation*}
    \frac{H}{R} = \frac{c_s}{v_k},
\end{equation*}

where $v_k$ is the keplerian velocity. Assuming a radial temperature profile $T \propto R^{-1/2}$, then $c_s \propto R^{-1/4}$ so that $H/R$ will scale as $R^{1/4}$. Note that this implies $\gamma = 1$, which will be the case from now on. We can parametrise the dependency on $M_{\star}$ as a power-law with exponent $\beta$:

\begin{equation}
    \frac{H}{R} = \frac{H}{R}\big|_{R=R_0} \left( \frac{R}{R_0} \right)^{1/4} \left( \frac{M_{\star}}{M_{\odot}} \right)^{\beta}.
    \label{eq:hoverr}
\end{equation}

\noindent In this work, we have considered three possible values for $\beta$ which correspond to different physical situations:

\begin{itemize}
    \item $H/R$ does not depend on the stellar mass, which implies $\beta = 0$;
    \item $c_s$ does not depend on the stellar mass, implying $\beta = -1/2$;
    \item $T \propto {M_{\star}}^{0.15}$ (derived by radiative transfer models - see \citealt{Sinclair2020}), which implies $\beta = -0.425$. Physically, this means that, while radiative transfer predicts that discs around lower mass stars should be colder due to their lower luminosity, this is a sub-dominant effect in setting the scale-height: the weaker gravity of these stars drives most of the change in scale-height.
\end{itemize}

Of these three possibilities, the third one is the more realistic; nonetheless, the resulting value of $\beta$ is very close to $- 1/2$ - meaning that we can approximate it to the second case, which makes mathematical calculations more straightforward. On the other hand, despite not being realistic, the first scenario is usually assumed for the sake of simplicity; for this reason, we decided to include it in this work. In conclusion, we considered $\beta$ to be either equal to $0$ or $1/2$.

\begin{figure*}
    \centering
    \includegraphics[width = 18cm, keepaspectratio]{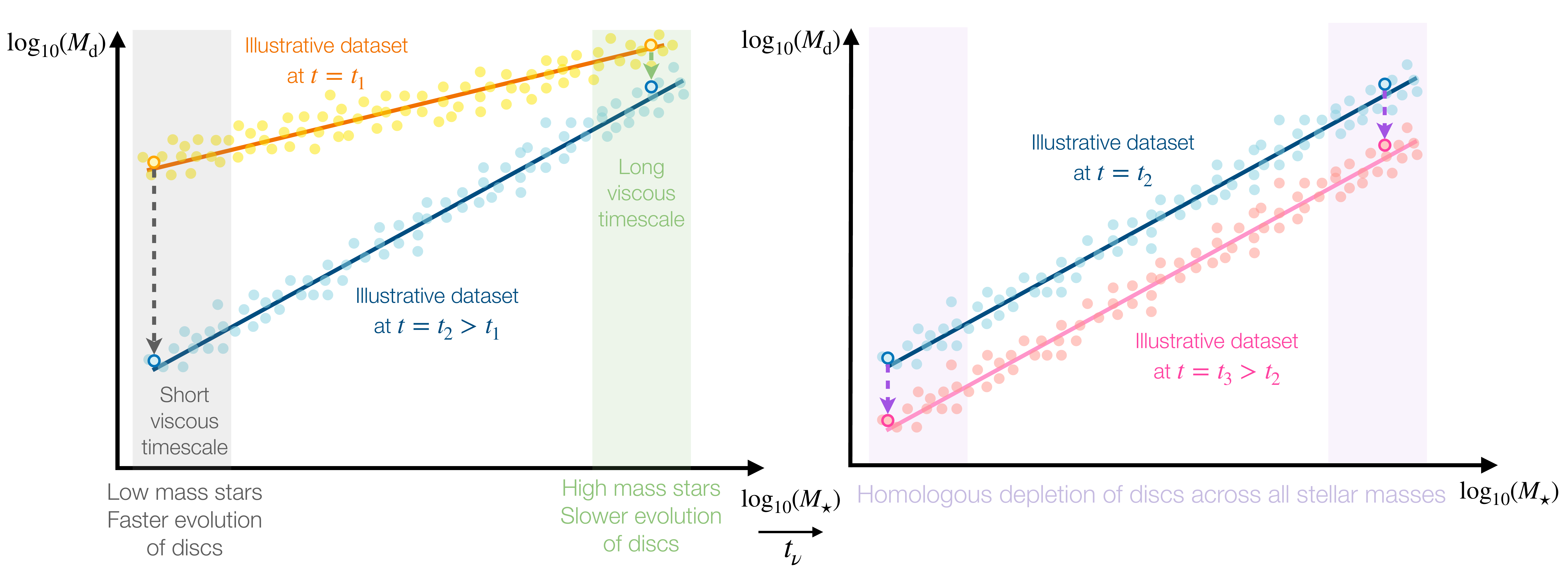}
    \caption{Illustrative sketch to show how steepening slopes are a consequence of a positive correlation between viscous timescales and stellar masses. The dots represent artificial datasets with their linear fits displayed as solid lines. The left panel shows an early evolutionary stage, where $t_{\nu} \propto {M_{\star}}^{\delta_0}$ with $\delta_0 > 0$: as discs around more massive stars have a longer viscous timescale, in the time interval $\Delta_t = t_2 - t_1$ their mass is less depleted than that of discs around less massive stars, which in turn have a short viscous timescale. This implies that, after $\Delta_t$, the correlation between $M_{\mathrm{d}}$ and $M_{\star}$ is steeper. On the other hand, the right panel represents a later evolutionary stage, where $\delta$ has reached the limit value of $0$. In that case, the viscous timescale does not depend on the stellar mass anymore and this implies a homologous depletion of discs around stars of all masses. The same argument applies for $\dot M$.}
    \label{fig:sketch}
\end{figure*}

\subsection{Young populations: \texorpdfstring{$t \ll t_{\nu}$}{tmintnu}} \label{subsec:young_pop}

If the age of the population is much smaller than its viscous timescale, no evolution has taken place yet. This means that we are still observing the initial condition, and that the parameters $\lambda_{\mathrm{m}}$, $\lambda_{\mathrm{acc}}$ and $\zeta$ coincide with $\lambda_{\mathrm{m}, 0}$, $\lambda_{\mathrm{acc}, 0}$ and $\zeta_0$. However, in the viscous scenario these parameters are always linked to one another; for young discs, Equation \eqref{eq:accrate_ss} can be written as $\dot M(0) \approx M_{\mathrm{d}}(0)/t_{\nu}$. In our notation, $M_\mathrm{d}$ represents the \textit{total} disc mass, which is given by $99 \%$ gas mass and $1 \%$ dust mass; since the observed power-law correlation (Equation \ref{eq:log_relation_mdiscmstar}) refers to the \textit{dust} mass, we need to translate $M_{\mathrm{dust}}$ to $M_\mathrm{d}$ dividing by the dust-to-gas ratio $\varepsilon$, which is assumed to be constant in time and $M_{\star}$. The initial, theoretical disc mass can therefore be written as

\begin{equation}
    M_{\mathrm{d}}(0) = K_1 \left( \frac{M_{\star}}{M_{\odot}} \right)^{\lambda_{\mathrm{m}, 0}},
    \label{eq:mdisc_zero_theo}
\end{equation}

\noindent where the normalization constant $K_1$ is given by $10^q M_{\oplus} / \varepsilon$. Taking into account the dependence of the aspect ratio of the disc $H/R$ on the stellar mass through Equation \eqref{eq:hoverr}, and using a generic value for $\beta$, we can write the initial accretion rate in the self-similar scenario as

\begin{equation}
    \dot M(0) = K_2 \left( \frac{M_{\star}}{M_{\odot}} \right)^{\lambda_{\mathrm{m}, 0} + 2 \beta + 1/2 - \zeta_0},
    \label{eq:mdot_zero_final}
\end{equation}

\noindent defining the normalization constant $K_2$ as

\begin{equation}
    K_2 = \frac{10^q M_{\oplus}}{\varepsilon} \sqrt{\frac{G M_{\odot}}{R_{c, \odot}}} \frac{3}{2} \alpha \left( \frac{H}{R}\big|_{R=R_{c, \odot}} \right)^2,
\end{equation}

\noindent where $R_{c, \odot}$ is the value of the cut-off radius $R_c$ for a solar-type star. The only free parameter in Equation \eqref{eq:mdot_zero_final} is the stellar mass $M_{\star}$: all of the other parameters ($\lambda_{\mathrm{m}, 0}$, $\zeta_0$, $R_{c, \odot}$, $\beta$) can be determined by comparing this expression to the observational data. In particular, since $\dot M \propto {M_{\star}}^{\lambda_{\mathrm{acc}, 0}}$, assuming that we can determine the values of $\lambda_{\mathrm{m}, 0}$ and $\lambda_{\mathrm{acc}, 0}$ from observations and that $\alpha$ is known we could constrain the value of $\zeta_0$,

\begin{equation}
    \zeta_0 \equiv \lambda_{\mathrm{m}, 0} + 2 \beta + \frac{1}{2} - \lambda_{\mathrm{acc}, 0}.
    \label{eq:zeta_initial}
\end{equation}

\noindent On the other hand, $R_{c, \odot}$ can be set imposing the normalization constant $K_2$ to equal the typical value for the accretion rate of a solar-type star. Based on the observation of \cite{Manara2017-Cha}, $\log_{10}(\dot M / M_{\odot} \text{yr}^{-1}) = -8.44$, which leads to $R_{c, \odot} = 1.6 \times 10^{14} \text{ cm} \approx 10 \text{ au}$ (assuming $\alpha = 10^{-3}$, as stated above). This is a reasonable initial disc radius; various recent papers have also assumed radii of the same order of magnitude \citep{Rosotti+Letter2019, Rosotti+2019, Trapman2019, Toci+2021}.

\subsection{Evolved populations: \texorpdfstring{$t \gg t_{\nu}$}{tbigtnu}} \label{subsec:evolved_pop}

If the age of the population is larger than its viscous timescale, discs can be considered evolved. In principle, at this stage the initial conditions are not observed anymore - meaning that we expect to see an evolution of both the slopes and spreads of the correlations. In particular, if $t \gg t_{\nu}$, the self-similar accretion rate reduces to

\begin{equation}
    \dot M (t) = \frac{1}{2} \frac{M_\mathrm{d}(t)}{t};
    \label{eq:mdot_final_evolved}
\end{equation}

\noindent as $t$ does not depend on the stellar mass, we can see that Equation \eqref{eq:mdot_final_evolved} implies that the dependency on $M_{\star}$ in evolved discs is the same for $\dot M$ and $M_\mathrm{d}$. Moreover, the distributions of the two quantities $M_{\mathrm{d}}$ and $\dot M$ are expected to resemble each other more and more as evolution takes place and the condition given in Equation \eqref{eq:mdot_final_evolved} is reached: this means that, eventually, we expect both the slope and the spread of the two correlations to reach the same value.

\subsubsection{Slopes}\label{subsub:evoslopes}

As we did in the young populations scenario, expressing $M_\mathrm{d}(t)$ through the first assumption in Equation \eqref{eq:assumptions} and using the definition of $t_{\nu}$ leads to

\begin{equation}
    M_\mathrm{d} (t) = K_3 \left( \frac{M_{\star}}{M_{\odot}} \right)^{\lambda_{\mathrm{m}, 0}+\zeta_0/2-\beta-1/4} t^{-1/2},
\end{equation}

\noindent where $K_3$ is a normalization constant that does not depend on the stellar mass or the age, defined by

\begin{equation*}
    K_3 = M_{\mathrm{d}, \odot} \left( \frac{{R_{c, \odot}}^3}{G M_{\odot}} \right)^{1/4} \frac{1}{\sqrt{3 \alpha}} \left( \frac{H}{R}\big|_{R=R_{c, \odot}} \right)^{-1};
\end{equation*}

\noindent $M_{\mathrm{d}, \odot}$ represents the initial disc mass for a solar-type star. Denoting the evolved slopes, at $t \gg t_{\nu}$, as $\lambda_{\mathrm{m}}$ and $\lambda_{\mathrm{acc}}$, this implies that

\begin{equation}
    \lambda_{\mathrm{m}} \equiv \lambda_{\mathrm{acc}} = \lambda_{\mathrm{m}, 0} + \frac{\zeta_0}{2} - \beta - \frac{1}{4}.
    \label{eq:lambda1e2_evolved}
\end{equation}

\noindent There are two ways in which Equation \eqref{eq:lambda1e2_evolved} can be satisfied:

\begin{itemize}
    \item $\lambda_{\mathrm{m}, 0} = \lambda_{\mathrm{acc}, 0}$: if the two slopes start from the same value, they do not evolve with time;
    \item $\lambda_{\mathrm{m}, 0} \neq \lambda_{\mathrm{acc}, 0}$: if the initial values of the two slopes are different, an evolution of their values must take place due to accretion processes.
\end{itemize}

In the first scenario, as we do not expect any evolution with time of the two slopes, we can use the initial condition (see Equation \ref{eq:zeta_initial}) to determine the value of $\zeta_0$, finding $\zeta_0 = 2 \beta + 1/2$. As expected, substituting this value of $\zeta_0$ in Equation \eqref{eq:lambda1e2_evolved} leads to $\lambda_{\mathrm{m}} = \lambda_{\mathrm{acc}} = \lambda_{\mathrm{m}, 0}$. This implies that observing the same slope for the two correlations $M_\mathrm{d} - M_{\star}$ and $\dot M - M_{\star}$ is not enough to claim anything on the age of the population. It could either be evolved, in which case we are observing a direct consequence of viscous evolution, or it could be young, and we would be observing the initial condition. This endorses what we suggested earlier, that observed correlations can be due either to the initial conditions or the evolution - or, most likely, a combination of the two. It would be possible to disentangle between the two possibilities if additional information was provided, for example

\begin{itemize}
    \item observations at earlier ages (to try and see a change in the slopes with time);
    \item measurements of disc radii $R_c$, which is in principle observable (although with some caveats - see \citealt{Toci+2021}) to determine the viscous timescale and give an estimate of the evolutionary stage of the population. Note that there is a degeneracy in the determination of the viscous timescale, as it depends on $\alpha$, $R_c$ and $H/R \big|_{R = R_c}$;
    \item the $\dot M - M_{\mathrm{d}}$ correlation: a large spread implies that the population cannot be considered evolved yet \citep{Lodato2017}.
\end{itemize}

The other possibility is that the two initial values of the slopes are different. If that is the case, we can define the difference between the initial values as $\delta_0 = \lambda_{\mathrm{m}, 0} - \lambda_{\mathrm{acc}, 0}$. This means that the initial condition for $\zeta_0$ (Equation \ref{eq:zeta_initial}) is given by $\zeta_0 = 1/2 + 2 \beta + \delta_0$, which can be used in Equation \eqref{eq:lambda1e2_evolved} to give

\begin{equation}
    \lambda_{\mathrm{m}} = \lambda_{\mathrm{acc}} = \frac{3 \lambda_{\mathrm{m}, 0} - \lambda_{\mathrm{acc}, 0}}{2}.
    \label{eq:lambda1e2_long}
\end{equation}

\noindent Note that Equation \eqref{eq:lambda1e2_long} is not symmetric in $\lambda_{\mathrm{m}, 0}$ and $\lambda_{\mathrm{acc}, 0}$, meaning that the impact of the initial condition for the $M_{\mathrm{d}} - M_{\star}$ correlation is more long-lived that that for $\dot M - M_{\star}$. Whether $\lambda_{\mathrm{m}, 0}$ and $\lambda_{\mathrm{acc}, 0}$ grow steeper or shallower with time, towards the limit value determined by Equation \eqref{eq:lambda1e2_long}, depends on the sign of $\delta_0$: 

\begin{itemize}
    \item A steepening of the correlation, $\lambda_{\mathrm{m}} > \lambda_{\mathrm{m}, 0}$, is achieved if $\lambda_{\mathrm{m}, 0} > \lambda_{\mathrm{acc}, 0}$ ($\delta_0 > 0$); this condition also implies $\lambda_{\mathrm{acc}} > \lambda_{\mathrm{acc}, 0}$;
    \item On the other hand, a flattening of the correlation, $\lambda_{\mathrm{m}} < \lambda_{\mathrm{m}, 0}$, is obtained if $\lambda_{\mathrm{m}, 0} < \lambda_{\mathrm{acc}, 0}$ ($\delta_0 < 0$); as in the previous case, this also implies $\lambda_{\mathrm{acc}} < \lambda_{\mathrm{acc}, 0}$.
\end{itemize}

\noindent In conclusion, there are two possible scenarios: either both slopes increase towards the common value for evolved populations, if $\lambda_{\mathrm{m}, 0} > \lambda_{\mathrm{acc}, 0}$, or they both decrease towards it, it $\lambda_{\mathrm{m}, 0} < \lambda_{\mathrm{acc}, 0}$. In both cases, the difference between the two slopes $\delta = \lambda_{\mathrm{m}} - \lambda_{\mathrm{acc}}$ will tend towards zero. This argument stems from purely mathematical considerations, but can be easily understood from the physical point of view. As the initial accretion timescale of the disc can be written as the ratio $M_{\mathrm{d}} / \dot M$, it will itself show a power-law correlation with the stellar mass. In particular, its slope will be the difference of those of $M_{\mathrm{d}} - M_{\star}$ and $\dot M - M_{\star}$; in our notation, its initial value is $\delta_0$. A positive correlation (i.e., a positive $\delta_0$) means that discs around more massive stars will evolve slower than discs around less massive stars, causing a steepening of the correlations between disc and stellar parameters. The contrary argument applies for a negative $\delta_0$, which leads to flattening correlations. As the evolution proceeds, $\delta = \lambda_{\mathrm{m}} - \lambda_{\mathrm{acc}} \rightarrow 0$, eventually reaching a homologous depletion of discs around more and less massive stars. Figure \ref{fig:sketch} illustrates this concept in form of sketch.

These conclusions on the evolutionary behaviour of $\lambda_{\mathrm{m}}$ and $\lambda_{\mathrm{acc}}$ only depend on the scaling of the viscous timescale with the stellar mass. This is the reason why our assumption that $\alpha$ does not depend on $M_{\star}$ does not influence this result. It should be noted, however, that if $\alpha$ did depend on the stellar mass we would find a different relation between $\lambda_{\mathrm{m}, 0}$, $\lambda_{\mathrm{acc}, 0}$ and $\zeta_0$, as the viscous timescale is set by $\alpha$ and $R_c$.

\subsubsection{Spread}\label{subsub:evospread}

As we have mentioned before, not only the mean slopes, but also the spreads of the $M_{\mathrm{d}} - M_{\star}$ and $\dot M - M_{\star}$ correlations are expected to reach the same limit value in the viscous scenario. \cite{Lodato2017} have shown that the $\dot M - M_{\mathrm{d}}$ correlation is expected to tighten in time, leading to a decreasing spread; on the other hand, as the stellar masses hardly change, we expect the $M_{\mathrm{d}} - M_{\star}$ correlation to follow the same behaviour of that of the distribution of disc masses, which we discuss below.

If we assume the initial disc masses and radii to be distributed log-normally, we can show that viscous evolution preserves the log-normal shape of the distributions. Moreover, the spread of the $M_{\mathrm{d}}$ distribution at $t \gg t_{\nu}$ writes (see Appendix \ref{appendix-spreads} for derivation)

\begin{equation}
    \sigma^2_M(t) = {\sigma^2_M}(0) + (1 - \eta)^2 {\sigma^2_{t_{\nu}}}(0);
    \label{eq:evospread_M}
\end{equation}

\noindent $\sigma_{t_{\nu}}$ is the spread of the distribution of viscous timescales, which in our case, given the direct proportionality between the viscous timescale and radius, is very close to that of radii $\sigma_R$. Equation \eqref{eq:evospread_M} shows that the dispersion in the distribution of disc masses increases with disc evolution.

Conversely, we expect the dispersion of the distribution of accretion rates to decrease with time. As the initial accretion rate can be written as $\dot M \propto M_{\mathrm{d}, 0}/t_{\nu}$, the initial dispersion $\sigma_{\dot M}(0)$ is given by ${\sigma^2_{\dot M}}(0) = {\sigma^2_{M}}(0) + {\sigma^2_{t_{\nu}}}(0)$. At evolved times, instead, we have already noticed that the two distributions will coincide: this means that also $\sigma_{\dot M}(t)$ will be given by Equation \eqref{eq:evospread_M}. For $\gamma = 1$, $(1 - \eta)^2 = 1/4$, meaning that $\sigma_{\dot M}(t) > \sigma_{\dot M}(0)$.

\subsection{Summary: implications of the different scenarios} \label{subsec:summary_implications}

If we assume a power-law dependence of the disc mass $M_\mathrm{d}$ and accretion rate $\dot M$ on the stellar mass $M_{\star}$ as initial condition for a population of discs, viscous theory predicts four different evolutionary scenarios for the observed correlations at later times:

\begin{itemize}
    \item young populations ($t \ll t_{\nu}$): evolution has not taken place yet, therefore the observed slopes $\lambda_{\mathrm{m}}$ and $\lambda_{\mathrm{acc}}$ are the same as the initial values $\lambda_{\mathrm{m}, 0}$ and $\lambda_{\mathrm{acc}, 0}$;
    \item old populations ($t \gg t_{\nu}$): evolution has taken place and both slopes have had the time to reach the evolved value, $\lambda_{\mathrm{m}} = \lambda_{\mathrm{acc}} = (3 \lambda_{\mathrm{m}, 0} - \lambda_{\mathrm{acc}, 0})/2$. This can take place either via
    \begin{enumerate}
        \item both slopes already starting at the final value $\lambda$, if $\lambda_{\mathrm{m}, 0} = \lambda_{\mathrm{acc}, 0}$ (implying $\delta_0 = \lambda_{\mathrm{m}, 0} - \lambda_{\mathrm{acc}, 0} = 0)$;
        \item a steepening of both slopes towards $\lambda_{\mathrm{m}} = \lambda_{\mathrm{acc}}$, if $\lambda_{\mathrm{m}, 0} > \lambda_{\mathrm{acc}, 0}$ (implying $\delta_0 > 0)$;
        \item a flattening of both slopes towards $\lambda_{\mathrm{m}} = \lambda_{\mathrm{acc}}$, if $\lambda_{\mathrm{m}, 0} < \lambda_{\mathrm{acc}, 0}$ (implying $\delta_0 < 0)$.
    \end{enumerate}
\end{itemize}

\begin{table}
    \centering
    \includegraphics[width = 6cm]{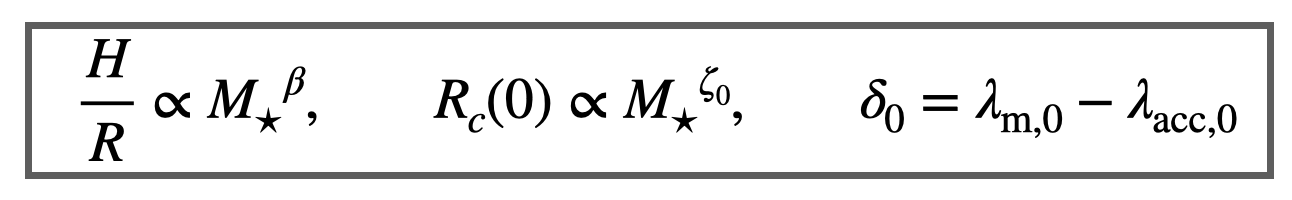} \\
    \vspace{0.3cm}
    \begin{tabular}[H]{P{1.3cm} P{1.3cm}}
        \multicolumn{2}{c}{ $\bm{\lambda_{\mathrm{m}, 0}} \mathbf{=} \bm{\lambda_{\mathrm{acc}, 0}} \Hquad \mathbf{\Rightarrow} \Hquad \bm{ \delta_0} \mathbf{= 0}$} \\[5pt]
        \hline
        \hline
        $\bm{\beta}$ & $\bm{\zeta_0}$ \\[3pt]
        \hline
        \cellcolor{gray!15} 0 & $\frac{1}{2}$ \\[2.5pt]
        \scalebox{1.2}{$- \frac{1}{2}$} & \cellcolor{gray!15} $- \frac{1}{2}$ \\[2.5pt]
    \end{tabular}

    \vspace{0.4cm}

    \begin{tabular}[H]{P{1.5cm} P{1.5cm} P{1.5cm}}
        \multicolumn{3}{c}{ $\bm{\lambda_{\mathrm{m}, 0}} \mathbf{>} \bm{\lambda_{\mathrm{acc}, 0}} \Hquad \Rightarrow \Hquad \bm{ \delta_0} \mathbf{> 0}$} \\[5pt]
        \hline
        \hline
        $\bm{\beta}$ & $\bm{\delta_0}$ & $\bm{\zeta_0}$ \\[3pt]
        \hline
        \cellcolor{gray!15} 0 & $\left(0, + \infty \right)$ & $\left[ \frac{1}{2}; + \infty \right)$ \\
        \scalebox{1.2}{$- \frac{1}{2}$} & $\left(0, \frac{1}{2} \right]$ & \cellcolor{gray!15} $\left( - \infty; 0 \right]$ \\
        \noalign {\global \arrayrulewidth = 0.6mm} \arrayrulecolor{YellowGreen}\hline
        \multicolumn{1}{|P{1.5cm}}{$- \frac{1}{2}$} & $\left[ \frac{1}{2}; + \infty \right)$ & \multicolumn{1}{P{1.5cm}|}{$\left[0; + \infty \right)$} \\
        \hline
    \end{tabular}

    \vspace{0.4cm}
    
    \begin{tabular}[H]{P{1.5cm} P{1.5cm} P{1.5cm}}
        \multicolumn{3}{c}{ $\bm{\lambda_{\mathrm{m}, 0}} \mathbf{<} \bm{\lambda_{\mathrm{acc}, 0}} \Hquad \mathbf{\Rightarrow} \Hquad \bm{ \delta_0} \mathbf{< 0}$} \\[5pt]
        \noalign {\global \arrayrulewidth = 0.2mm}\hline
        \hline
        $\bm{\beta}$ & $\bm{\delta_0}$ & $\bm{\zeta_0}$ \\[3pt]
        \hline
        \cellcolor{gray!15} 0 & $\left( - \infty; - \frac{1}{2} \right]$ & \cellcolor{gray!15} $\left( - \infty; 0 \right]$ \\
        \cellcolor{gray!15} 0 & $\left[ - \frac{1}{2}; 0 \right)$ & $\left[ 0; \frac{1}{2} \right]$ \\
        \scalebox{1.2}{$- \frac{1}{2}$} & $\left(- \infty; 0 \right)$ & \cellcolor{gray!15} $\left(- \infty; - \frac{1}{2} \right]$ \\
    \end{tabular}
    
    \vspace{1cm}
    \caption{Summary of the different possible theoretical scenarios divided by the relative values of $\lambda_{\mathrm{m}, 0}$ and $\lambda_{\mathrm{acc}, 0}$ (see text for details, especially Equation \ref{eq:zeta_initial} and paragraph \ref{subsec:summary_implications}). The definition of each parameter appearing in the table is reminded in the top grey box. Each line represents a set of parameters. As discussed in the text, we dismiss the cases where $\beta = 0$ (which would imply an aspect ratio independent on $M_{\star}$) and $\zeta_0 < 0$ (implying disc radii decreasing with $M_{\star}$): for visual purposes, cells corresponding to these two conditions have a gray background. The only line which does not include any gray cells is the one that is not influenced by any of these restraints, i.e., the most physically reasonable scenario, and is framed in green.}
    \label{tab:table_results}
\end{table}

Table \ref{tab:table_results} summarises all of the possible theoretical scenarios, each line representing a set of parameters. The steps to determine a set of values are as follows:

\begin{enumerate}
    \item first, the initial values of the slopes of the correlations $M_\mathrm{d} - M_{\star}$ and $\dot M - M_{\star}$, $\lambda_{\mathrm{m}, 0}$ and $\lambda_{\mathrm{acc}, 0}$, are chosen. This determines $\delta_0$, defined as the difference between the two;
    \item $\beta$, the slope of the correlation $H/R - M_{\star}$ (Equation \ref{eq:hoverr}), is chosen;
    \item using Equation \eqref{eq:zeta_initial}, $\zeta_0$ - the slope of the correlation $R_{\mathrm{d}} - M_{\star}$ - is determined. 
\end{enumerate}

However, as we discussed above, not all parameters can assume any possible theoretical value while still maintaining physical relevance. In particular, $\beta$ (the slope of the correlation of $H/R$ with $M_{\star}$) is unlikely to be zero, and $\zeta_0$ (the initial slope of the correlation between $R_c$ and $M_{\star}$) is unlikely to be negative. These conditions are visualized in Table \ref{tab:table_results} as cells with a gray background. Note that the assumption of a constant $\alpha$ does influence this argument: the increasing, decreasing and constant behaviour of $\lambda_{\mathrm{m}}$ and $\lambda_{\mathrm{acc}}$ based on the initial conditions is not affected, but the physical likelihood of the different scenarios is. Some lines in Table \ref{tab:table_results} show the same value of $\beta$ but different intervals for $\delta_0$; this is because based on the value of $\delta_0$, different possibilities for $\zeta_0$ may arise - in particular, it is meaningful to split the different possibilities if they include a change in the sign of $\zeta_0$. Given that each line in the table represents a set of parameters, lines that contain one or more gray cells turn out to be discarded.

The only line that is not influenced by these restrictions, framed in green, represents the physically meaningful scenario. It is interesting to note that the requirement for both $\beta$ and $\zeta_0$ to assume reasonable values determines the difference between the initial slopes $\lambda_{\mathrm{m}, 0}$ and $\lambda_{\mathrm{acc}, 0}$. Specifically, it prescribes $\delta_0$ to be greater than $1/2$: this implies $\lambda_{\mathrm{m}, 0} > \lambda_{\mathrm{acc}, 0}$, making the steepening slopes scenario the most reasonable one. Moreover, as we discussed in the previous paragraph, $\delta_0 > 0$ also implies a positive initial correlation between the accretion timescale and the stellar mass. The observational claim by \citetalias{Ansdell+2017} is intriguingly matching our theoretical consideration; a preliminary comparison with data and a discussion of its implications is performed in subsection \ref{subsec:obsdata}.

\begin{figure*}
    \begin{minipage}[t]{0.333\textwidth}
        \includegraphics[width=\linewidth]{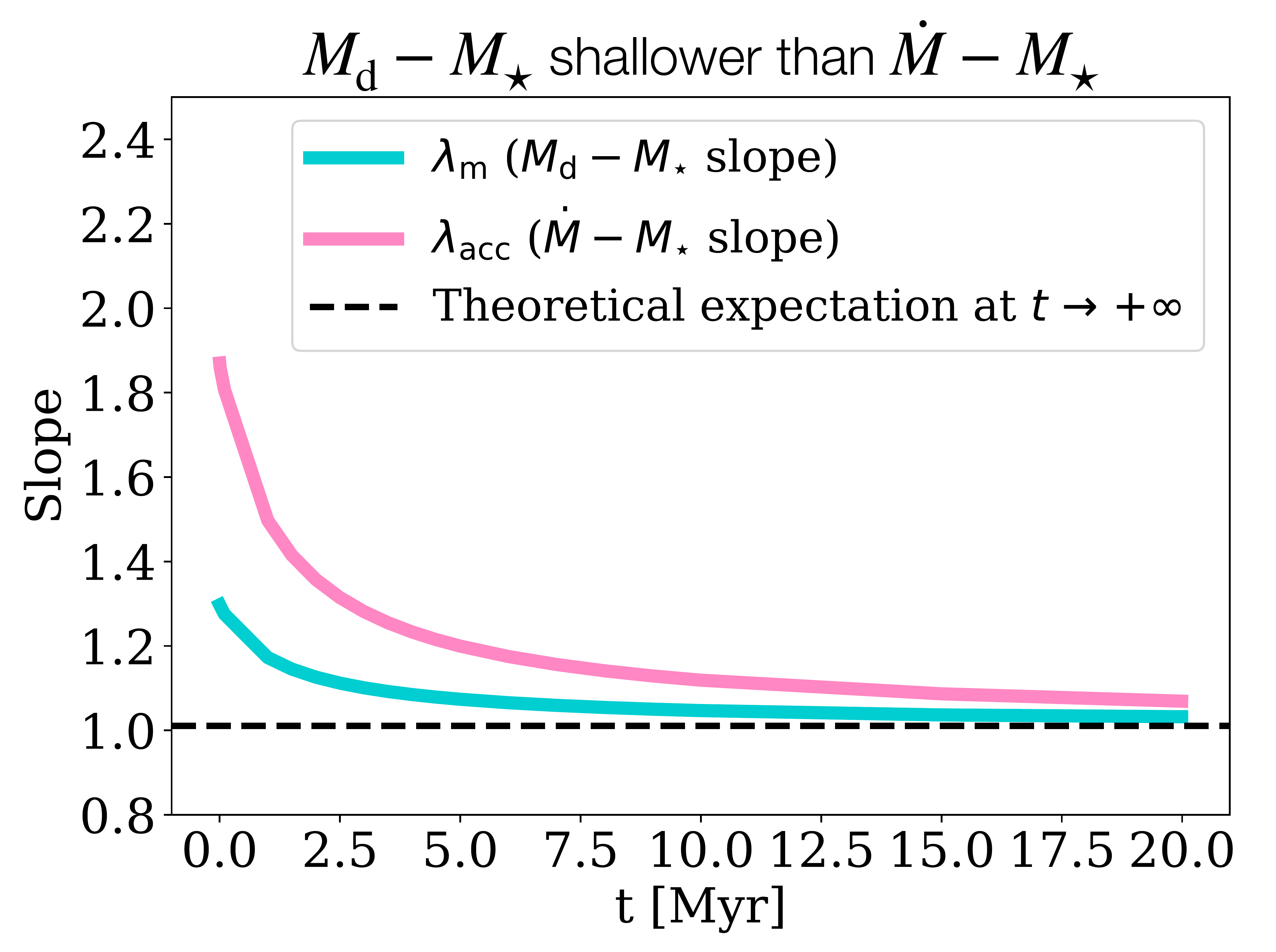}
    \end{minipage}%
    \hfill
    \begin{minipage}[t]{0.333\textwidth}
        \includegraphics[width=\linewidth]{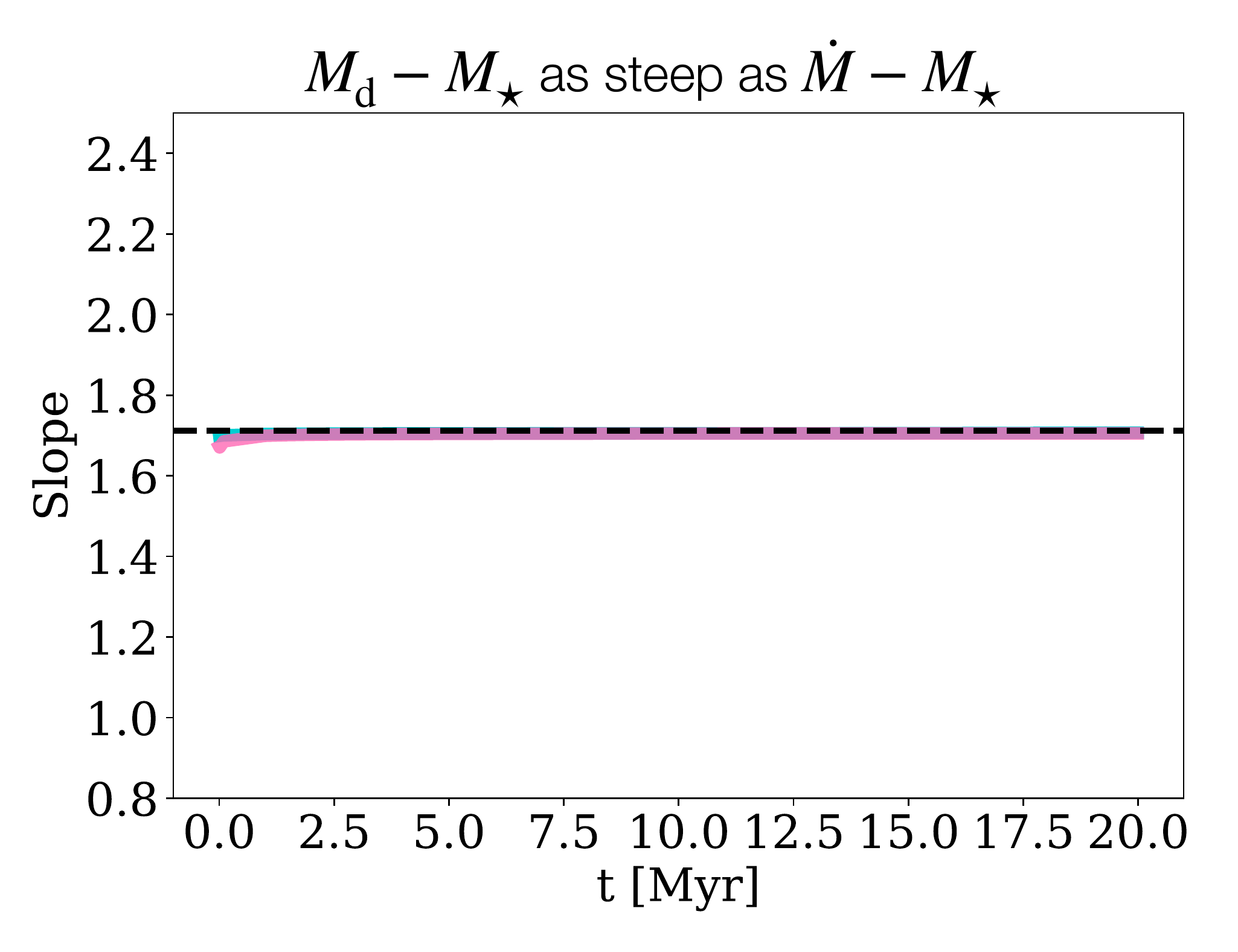}
        \end{minipage}%
    \hfill
    \begin{minipage}[t]{0.333\textwidth}
        \includegraphics[width=\linewidth]{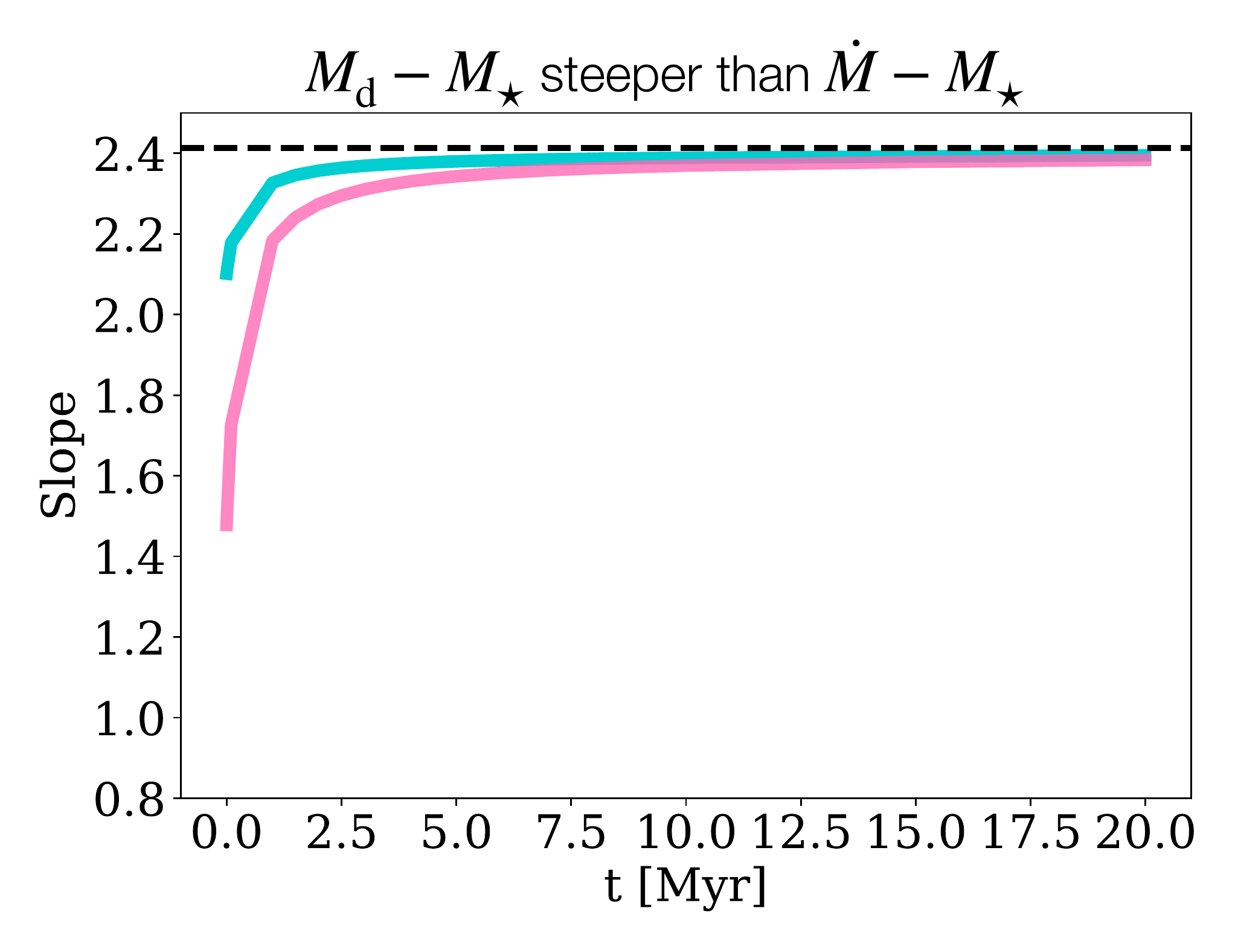}
    \end{minipage}
    \caption{Time evolution of the mean values of the slopes of the $M_\mathrm{d} - M_{\star}$ and $\dot M - M_{\star}$ correlations, $\lambda_{\mathrm{m}}$ and $\lambda_{\mathrm{acc}}$ (cyan and pink respectively), as obtained running \texttt{Diskpop} and fitting the results using \texttt{linmix}. The black dashed line represents the limit value of both slopes for $t \to + \infty$ (see Equation \ref{eq:lambda1e2_long}). Each panel corresponds to a different choice in the initial values $\lambda_{\mathrm{m}, 0}$ and $\lambda_{\mathrm{acc}, 0}$. Left panel: $\lambda_{\mathrm{m}, 0} = 1.3$, $\lambda_{\mathrm{acc}, 0} = 1.9$, $\delta_0 = \lambda_{\mathrm{m}, 0} - \lambda_{\mathrm{acc}, 0} < 0$; middle panel: $\lambda_{\mathrm{m}, 0} = \lambda_{\mathrm{acc}, 0} = 1.7$, $\delta_0 = 0$; right panel: $\lambda_{\mathrm{m}, 0} = 2.1$, $\lambda_{\mathrm{acc}, 0} = 1.5$, $\delta_0 >0$. The numerical behaviour follows the theoretical prediction derived in Section \ref{sec:analytical_considerations}: $\lambda_{\mathrm{m}, 0} < \lambda_{\mathrm{acc}, 0}$ (left panel) leads to a flattening trend, $\lambda_{\mathrm{m}, 0} = \lambda_{\mathrm{acc}, 0}$ (middle panel) to a constant slope, and $\lambda_{\mathrm{m}, 0} > \lambda_{\mathrm{acc}, 0}$ to a steepening trend. Moreover, the theoretical limit value is recovered for every choice of parameters. In every simulation, $H/R\big|_{R= 1 \text{ au}} = 0.33$ for $M_{\star} = 1 M_{\odot}$, $\alpha = 10^{-3}$, $\beta = - 0.5$, and both initial conditions $M_{\mathrm{d}}(0)$ and $R_c(0)$ follow a log-normal distribution.}
    \label{fig:diskpop_results}
\end{figure*}

\section{Population synthesis} \label{sec:population_synthesis}

In this Section we discuss the numerical counterpart of the theoretical analysis presented in Section \ref{sec:analytical_considerations}. In particular, we describe the population synthesis model that we have implemented in \texttt{Diskpop}: we briefly present the physical framework, and we show the numerical results as well as discussing their agreement with the theoretical predictions.

\subsection{Numerical methods - \texttt{Diskpop}} \label{subsec:diskpop}

Developing a population synthesis model requires to implement a numerical code to generate and evolve a synthetic population of protoplanetary discs. In this paper, we have used the Python code \texttt{Diskpop} that we developed and that we will release soon. In this paragraph, we briefly describe its basic functioning; for a more detailed presentation of it, we refer to the upcoming paper.

\texttt{Diskpop} generates an ensemble (which we term \textit{population}) of $N$ discs using the following scheme:

\begin{enumerate}
    \item determines $N$ stellar masses: this is achieved performing a random draw from an input probability distribution. In \texttt{Diskpop}, we use the initial mass function proposed by \cite{Kroupa2001};
    \item determines the $N$ mean values of initial disc masses and radii: this is where we account for correlations between stellar and disc parameters. Following the initial correlations listed in Equation \eqref{eq:assumptions}, we choose the values of $\lambda_{\mathrm{m}, 0}$ and $\lambda_{\mathrm{acc}, 0}$ and evaluate $\zeta_0$ using Equation \eqref{eq:zeta_initial}; the initial mean disc mass and radius are then computed, for each of the $N$ stellar masses from step (i), using the prescriptions \eqref{eq:mdisc_zero_theo} and \eqref{eq:mdot_zero_final}. Note that, by doing so, the correlations are intrinsically set to hold only for the mean value of the relevant parameters;
    \item draws the $N$ disc masses and radii: for this, the user needs to set two distributions (usually normal or log-normal) and their initial spread in dex (usually between 0.5 and 1.5 dex). After these parameters are determined, the draw is performed using the mean values computed in step (ii).
\end{enumerate}

The other relevant parameters besides $M_{\star}$, $M_\mathrm{d}$ and $R$ are fixed in our model. In particular, the aspect ratio of the disc at $1$ au and the $\alpha$ parameter for the \cite{ShakuraSunyaev1973} prescription are set to be $\frac{H}{R} \big|_{R = 1 \text{ au}} = 0.03$ for a 1 $M_{\odot}$ star and $\alpha = 10^{-3}$. As for the value of $\alpha$, it is still very much debated how high, or low, it should be or even if it should be constant in time or across the discs in a population at all \citep{Rafikov2017}; some works have also suggested a dependence of $\alpha$ on the radial position in the disc, namely a lower value in the inner disc with respect to the outer disc \citep{Liu+2018AlphaVarying}. Nonetheless, assuming a constant value of around $10^{-3}$ leads to reproducing the observed evolution. A proper study of the $\alpha$ value is out of the scope of this paper.

Once the population is initialized, we want to evolve it using the chosen prescription. In the purely viscous case, which is the focus of this first paper, we can either use the analytical self-similar solution \eqref{eq:selfsimilarsol} or solve the viscous evolution equation,

\begin{equation}
    \frac{\partial \Sigma}{\partial t} = \frac{3}{R} \frac{\partial}{\partial R} \left( R^{1/2} \frac{\partial}{\partial R} (\nu \Sigma R^{1/2}) \right);
    \label{eq:viscousevo}
\end{equation}

\noindent to do so, \texttt{Diskpop} employs the Python code presented in \cite{Booth+2017-viscouscode}. In this work we have used the numerical approach, but as long as an analytical solution is available there is no difference in the two methods.

The raw numerical solution is an array of values for the gas surface densities $\Sigma$ at the chosen ages, from which we compute the other quantities of interest: the disc mass is defined as the integral of the surface density from the inner ($R_{in}$) to the outer ($R_{out}$) radius,

\begin{equation}
    M_{\mathrm{d}} = \int_{R_{in}}^{R_{out}} 2 \pi r \Sigma \,dr,
    \label{eq:discmass}
\end{equation}

\noindent and $R$ as the radius enclosing 68 $\%$ of the total disc mass.

\subsection{Numerical simulations and comparison with theoretical expectations}\label{subsec:numericalsim_and_comparison}

In this section, we show the results obtained from simulating populations of $N = 100$ protoplanetary discs using \texttt{Diskpop}, evolved via numerical solution of Equation \eqref{eq:viscousevo}. Following the scheme of Section \ref{sec:analytical_considerations}, we divide the following discussion on the basis of the evolutionary regime considered. Our aim is to compare population synthesis results with the analytical prescription that we derived in the previous Section.

\subsubsection{Young populations: \texorpdfstring{$t \ll t_{\nu}$}{tmintnu}} \label{subsubsec:numerical_young_pop}

The discussion in section \ref{subsec:young_pop} shows that, if the considered population is much younger than its viscous timescale, we do not expect any significant evolution of the initial conditions. It is worth making a couple of considerations on the order of magnitude of the viscous timescale. For $\gamma = 1$, $t_{\nu}$ scales with the disc radius (as well as $\alpha^{-1}$, which is fixed in this work though), and given that our initial disc sizes are of the order of $10$ au, the typical viscous timescale will be shorter than $1$ Myr; this is the case for more than $98 \%$ of the discs in the synthetic population. This constraint stems from the need to reproduce the mean observed mass and accretion rate, as discussed in section \ref{subsec:young_pop}. Such a short viscous timescale means that young populations would be very challenging to observe. The earliest data available for populations of discs usually correspond to ages around 1 Myr (e.g., Lupus and $\rho$ Ophiucus); if the distribution of $t_{\nu}$ reproduced the viscous timescales of these regions, already the youngest populations would be too old for $t \ll t_{\nu}$ to hold. For this reason, in the following we only show numerical results for the evolved populations scenario. As mentioned earlier, all of this discussion holds for the particular viscous timescales that derive from the chosen parameters in our simulations: there is in principle the possibility of obtaining longer viscous timescales (lower $\alpha$, larger $R_c$), which would reverse this argument. However, from the observational point of view, this would require larger disc masses or lower accretion rates.

\subsubsection{Evolved populations: \texorpdfstring{$t \gg t_{\nu}$}{tbigtnu}} \label{subsubsec:numerical_evolved_pop}

In the case of evolved populations, which are observed at ages longer than their viscous timescales, we present the numerical results for all of the three scenarios discussed in section \ref{subsec:summary_implications}. 
Figure \ref{fig:diskpop_results} shows the mean fitted slopes of the $M_\mathrm{d} - M_{\star}$ and $\dot M - M_{\star}$ correlations ($\lambda_{\mathrm{m}}$ and $\lambda_{\mathrm{acc}}$ respectively) versus the age of the population. As summarised in section \ref{subsec:summary_implications}, from the theoretical point of view we expect both $\lambda_{\mathrm{m}}$ and $\lambda_{\mathrm{acc}}$ to reach the same evolved value, either through a steepening, a flattening or a constant evolution. Each of these three possibilities correspond to a choice of the initial parameters, and is determined by the sign of their difference $\delta_0 = \lambda_{\mathrm{m}, 0} - \lambda_{\mathrm{acc}, 0}$. In Figure \ref{fig:diskpop_results} we present our results, where the cyan and pink lines refer to $\lambda_{\mathrm{m}}$ and $\lambda_{\mathrm{acc}}$ respectively, obtained in each of these scenarios:

\begin{itemize}
    \item the left panel corresponds to $\lambda_{\mathrm{m}, 0} = 1.3, \Hquad \lambda_{\mathrm{acc}, 0} = 1.9 \Rightarrow \delta_0 < 0$, which is expected to produce a flattening of the slopes towards the evolved value;
    \item the middle panel shows the evolution if $\lambda_{\mathrm{m}, 0} = \lambda_{\mathrm{acc}, 0} = 1.7 \Rightarrow \delta_0 = 0$, which corresponds to the theoretical expectation of constant slopes;
    \item finally, in the right panel we set $\lambda_{\mathrm{m}, 0} = 2.1, \Hquad \lambda_{\mathrm{acc}, 0} = 1.5 \Rightarrow \delta_0 >0$, which should lead to an increasing trend of the slopes.
\end{itemize}

All simulations have the same values of $\alpha = 10^{-3}$, $H/R \big|_{R = 1 \text{au}} = 0.03$ for $M_{\star} = 1 M_{\odot}$, $\beta = - 1/2$. The black dashed line shows the theoretical evolved value, computed based on the initial values of the slopes as per Equation \eqref{eq:lambda1e2_long}. We can see that all of the three simulations do match our analytical evolutionary predictions: the limit value for evolved populations is recovered, and the flattening, constant and increasing trends are recovered. The middle panel shows a small evolution at very early ages, but it is most likely due to numerical effects and can be neglected. Note that the synthetic populations showed in the plots are evolved up to $20$ Myrs; while we do not expect to observe disc-bearing protostars at those very late ages (due to the disc removal processes at play that this work does not include - see \citealt{Mamajek2009}), we evolved our synthetic populations up to such a long age to make sure no transient affected our results. Moreover, the evolved value is reached at very late ages ($\sim 10$ Myrs), when we expect processes other than viscous evolution to have taken place.

\subsection{Numerical simulations introducing a spread}\label{subsec:sumericalsim_with_spread}

\begin{figure}
    \begin{minipage}[b]{0.45\linewidth}
        \centering
        \includegraphics[width=3.45cm, keepaspectratio]{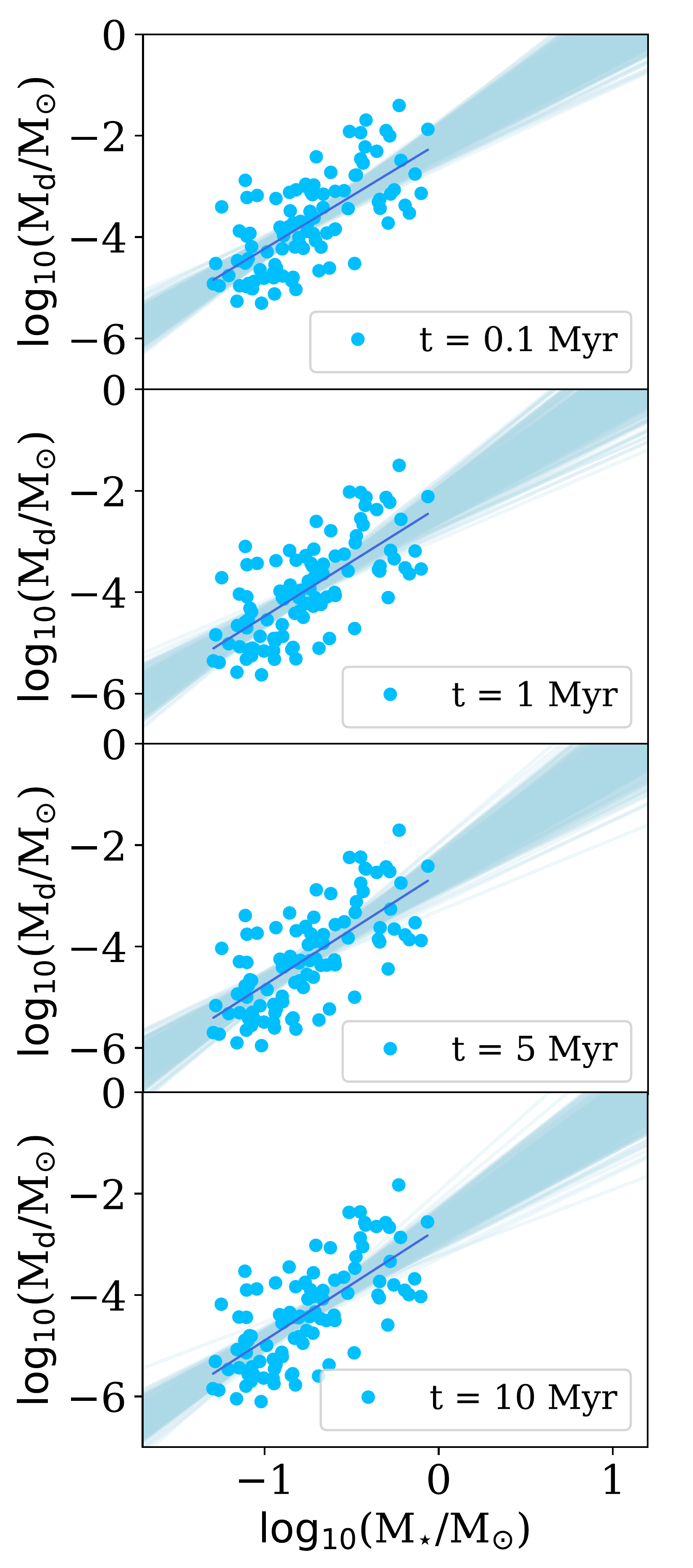}
    \end{minipage}
    \begin{minipage}[b]{0.45\linewidth}
        \centering
        \includegraphics[width=3.8cm, keepaspectratio]{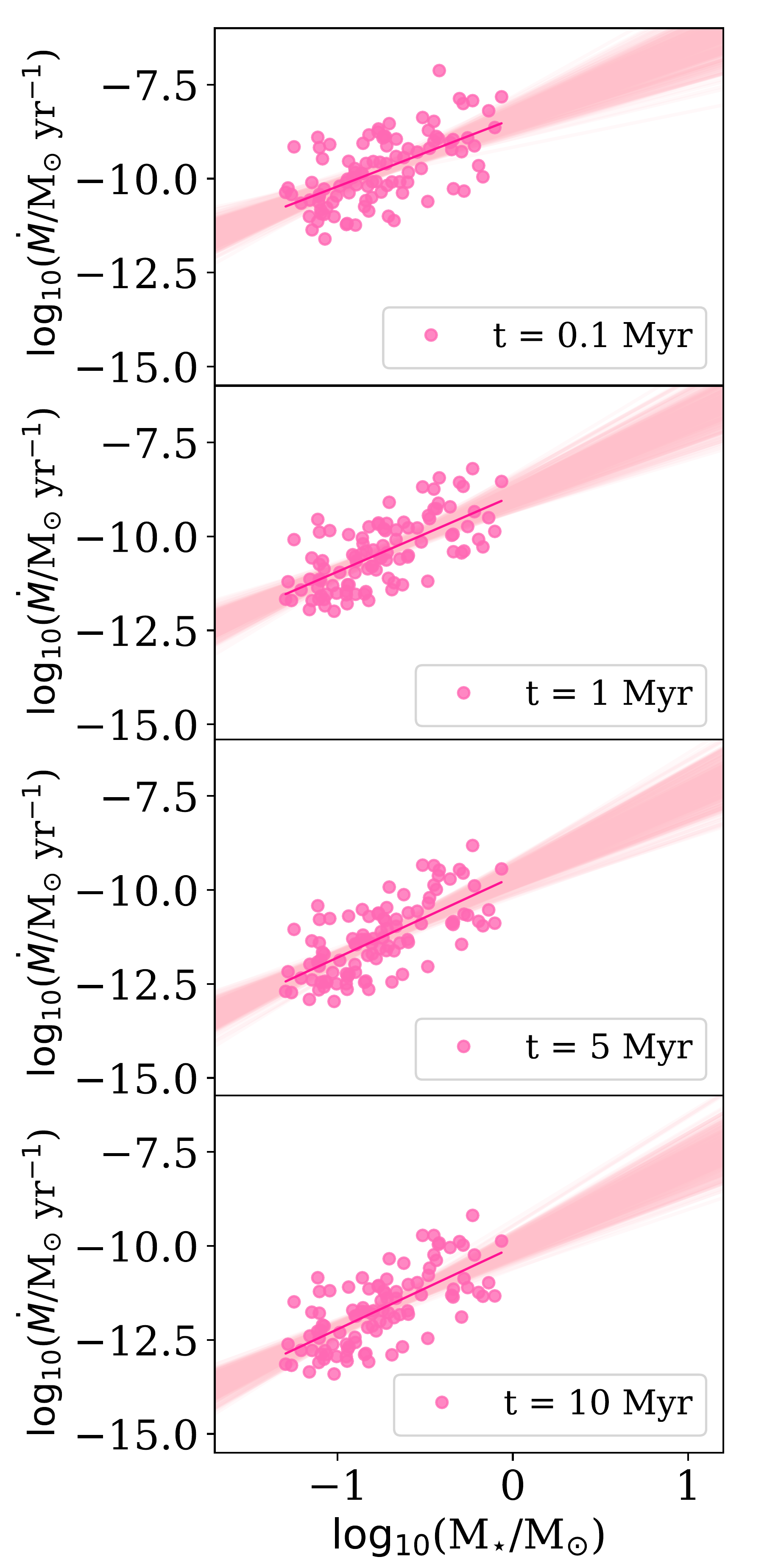}
    \end{minipage}
    \caption{$M_\mathrm{d} - M_{\star}$ (left panel) and $\dot M - M_{\star}$ (right panel) correlations obtained with \texttt{Diskpop} at four consequent time steps from top to bottom, as shown in the legend. The dots represent discs in the population, and the numerical fit is overplotted. These simulations were performed with the same parameters as the right panel of Figure \ref{fig:diskpop_results} ($\delta_0 > 0$), with the addition of a spread in the initial conditions of $\sigma_{M, 0} = 0.65$ dex and $\sigma_{R, 0} = 0.52$ dex.}
    \label{fig:showpop}
\end{figure}

\begin{figure}
    \centering
    \includegraphics[width = 7cm, keepaspectratio]{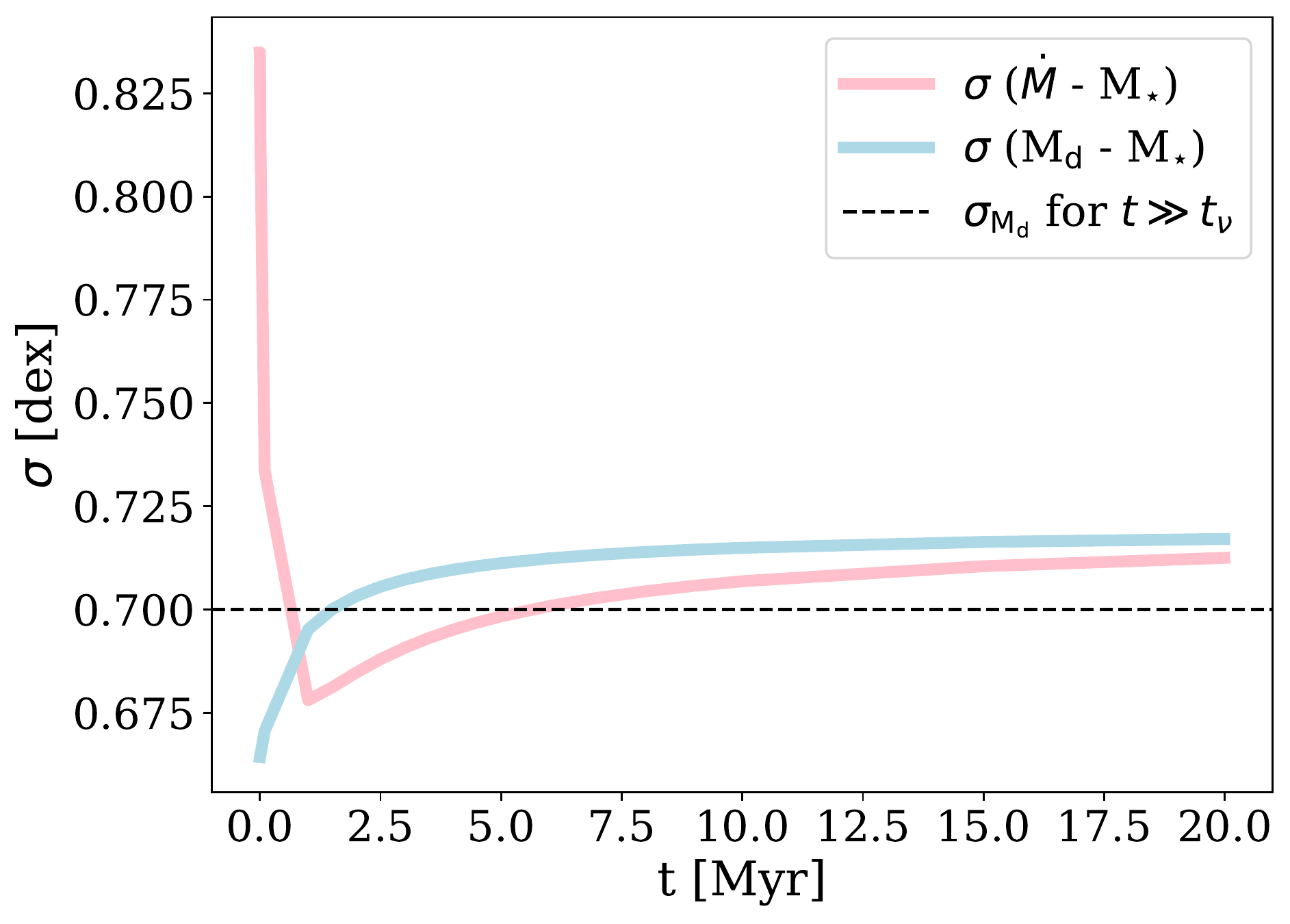}
    \caption{Time evolution of the spread of the $M_\mathrm{d} - M_{\star}$ (light blue) and $\dot M - M_{\star}$ (pink) correlations. The black dashed line shows the analytical spread of the $M_{\mathrm{d}}$ distribution for $t \gg t_{\nu}$ as per Equation \eqref{eq:evospread_M}. Both spreads reach the same value after some Myrs of evolution, in agreement with $M_\mathrm{d}$ and $\dot M$ reaching a unity correlation in the logarithmic plane; this final spread slightly differs from the analytical estimation for evolved populations due to the addition of a spread in the stellar masses.}
    \label{fig:sigmaevo}
\end{figure}

The real power of disc population syntheses is the possibility of introducing a spread in the initial conditions of the population: in this paragraph, we explore the influence of such a spread on our results. Figure \ref{fig:showpop} shows the output of \texttt{Diskpop}: the left and right panels display the correlation $M_\mathrm{d} - M_{\star}$ and $\dot M - M_{\star}$ respectively, at four subsequent timesteps as per the legend. Each dot represents a disc in the population; the fit was performed using the Python package \texttt{linmix}.

\begin{figure}
    \centering
    \begin{subfigure}[b]{0.48\textwidth}
       \includegraphics[width=1\linewidth]{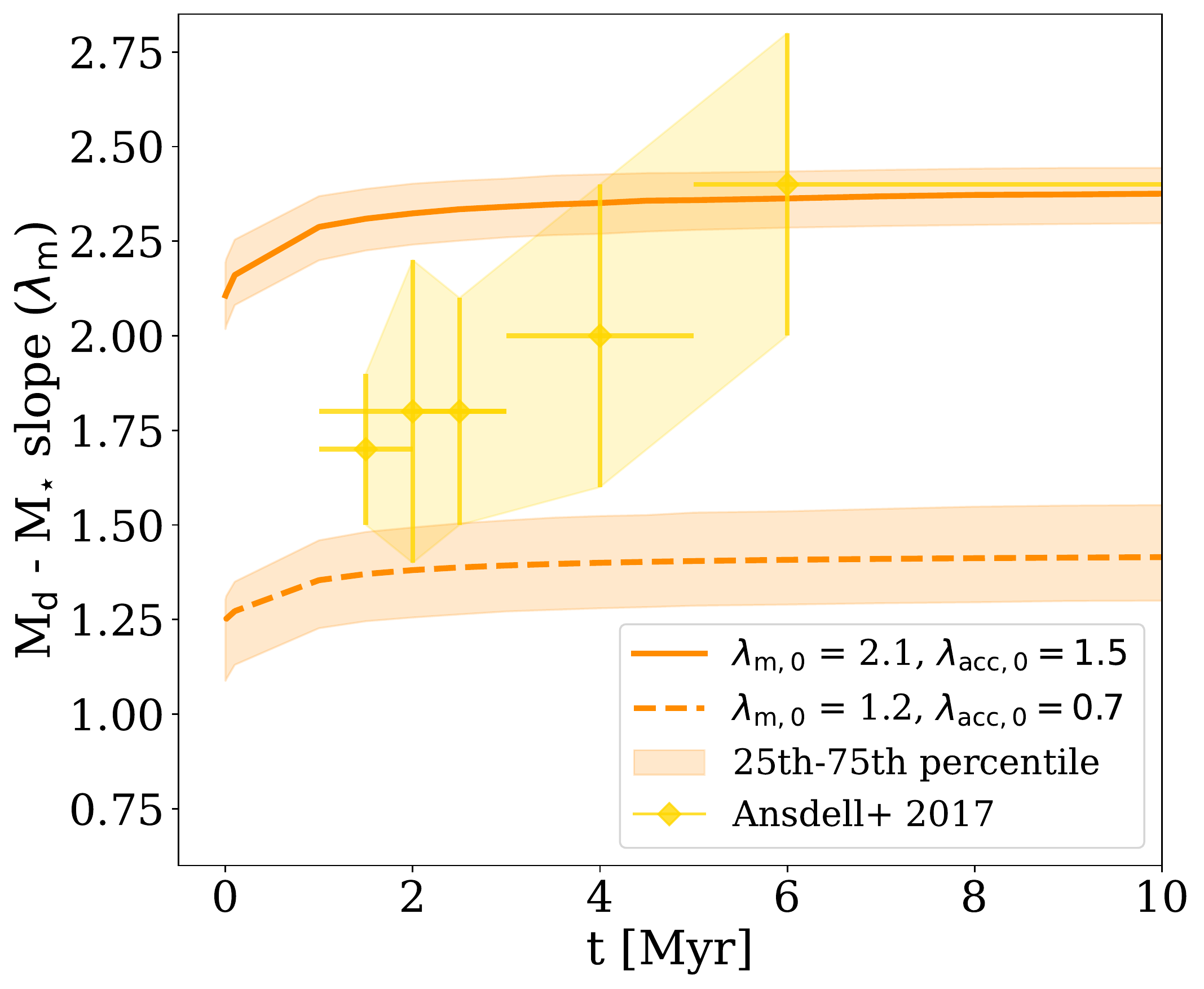}
    \end{subfigure}

\begin{subfigure}[b]{0.48\textwidth}
   \includegraphics[width=1\linewidth]{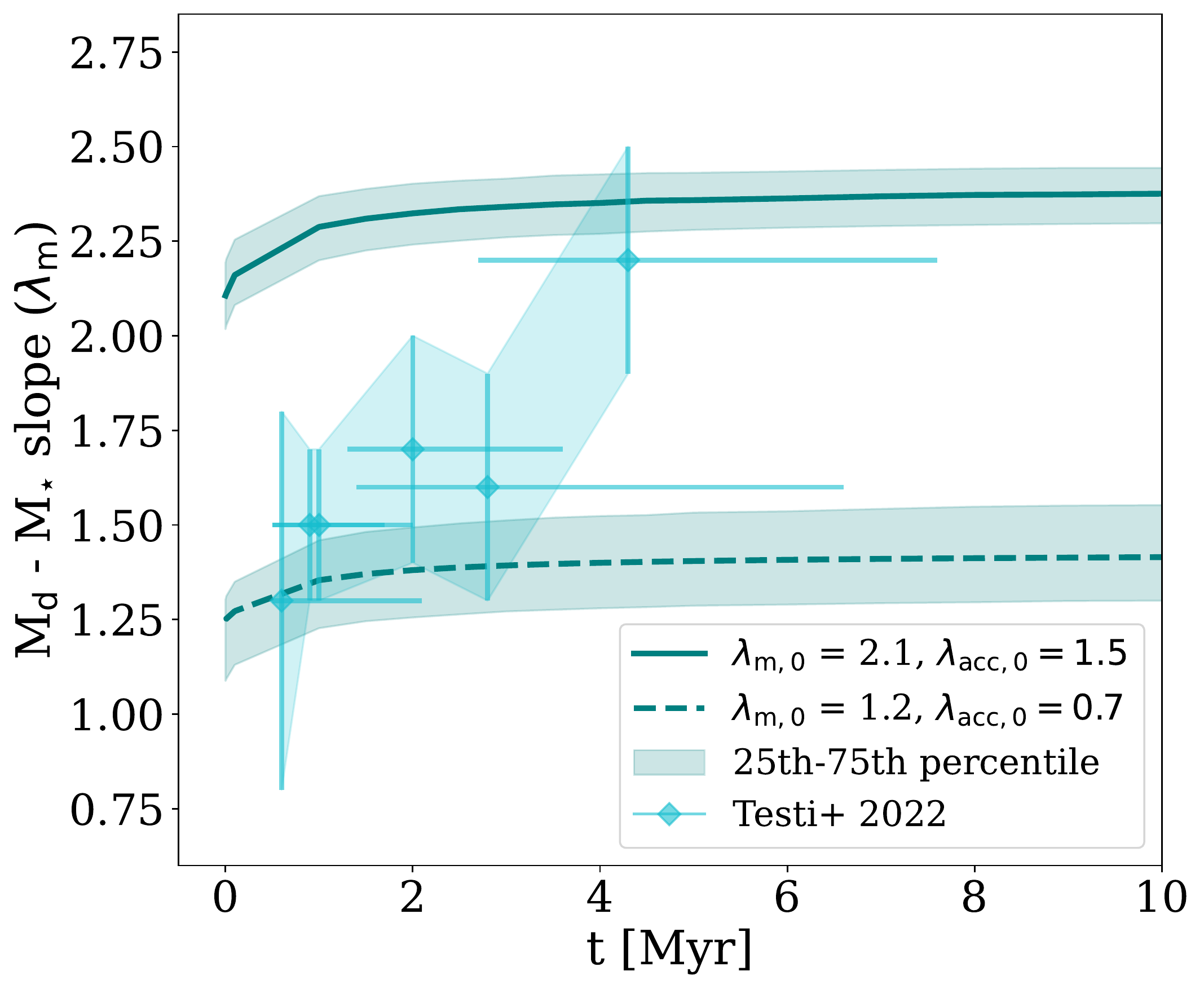}
\end{subfigure}

\caption{Comparison of the numerical results for $\lambda_{\mathrm{m}}$ obtained from \texttt{Diskpop} with the observational data from \protect\citetalias{Ansdell+2017} (top panel) and \protect\citetalias{Testi+2022-Ofiucone} (bottom panel). The solid and dashed lines in each panel represent the results of simulations performed with two different sets of parameters, as shown in the legend. The diamonds correspond to the observational mean values, while the vertical and horizontal lines represent the $1 \sigma$ error bars. The two sets of parameters displayed are able to match (from a qualitative point of view) the youngest and oldest populations, respectively; they represent the limits of the parameter space. Both are chosen to fall in the most physically meaningful scenario, with $\delta > 1/2$. The darker shaded regions around the solid and dashed lines represent the 25th and 75th percentile for the fitted slopes, obtained from 100 different statistical realisations of the same simulation.}
\label{fig:data_comp_mdisc}
\end{figure}

\begin{figure}
    \centering
    \includegraphics[width = 8.5cm]{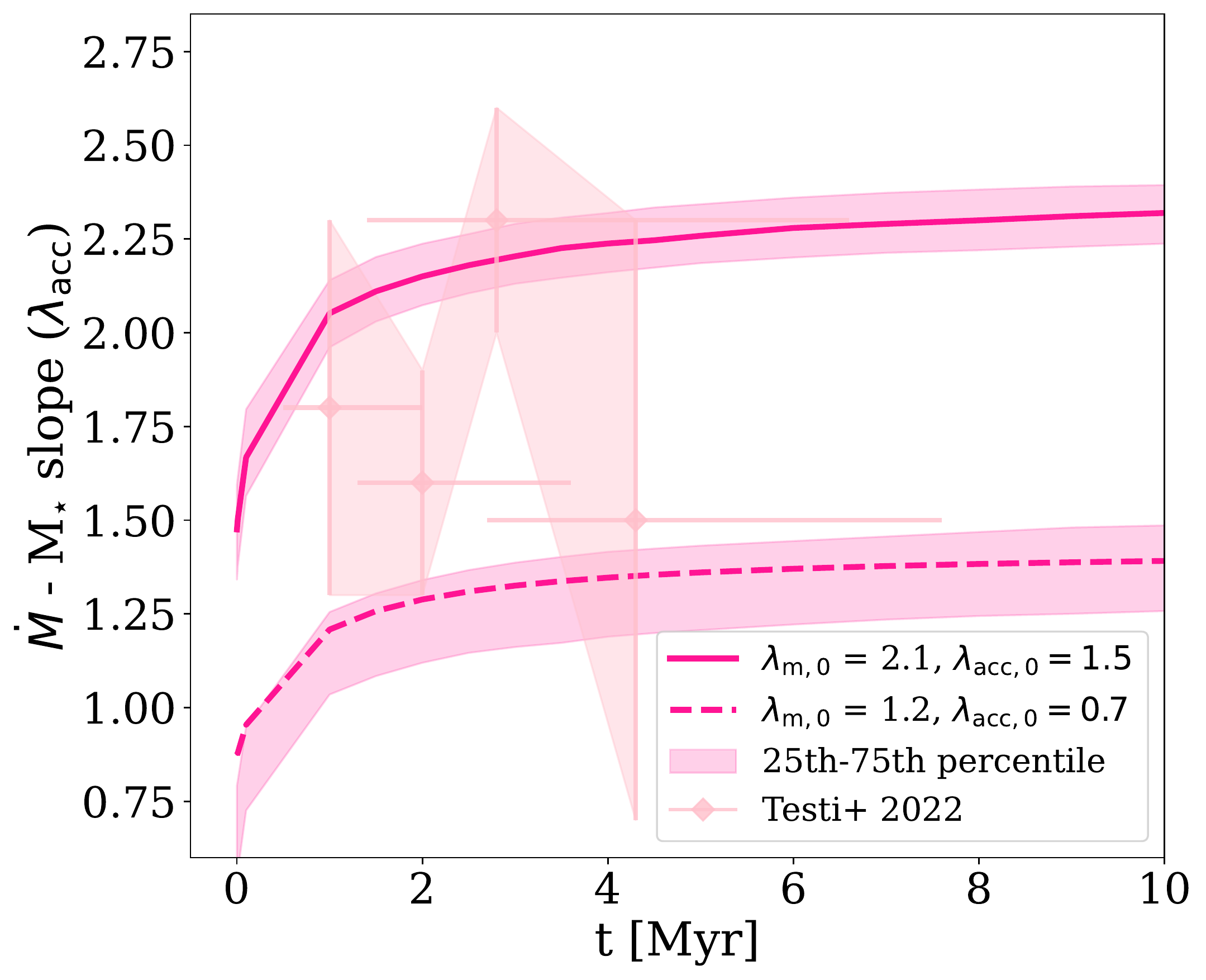}
    \caption{Same as Figure \ref{fig:data_comp_mdisc}, but referring to $\lambda_{\mathrm{acc}}$, showing both the numerical results from \texttt{Diskpop} and the observational data from \protect\citetalias{Testi+2022-Ofiucone}. See the caption of Figure \ref{fig:data_comp_mdisc} for detail on annotations.}
    \label{fig:data_comp_mdot_Testi}
\end{figure}

\begin{figure}
    \centering
    \includegraphics[width = 8.5cm]{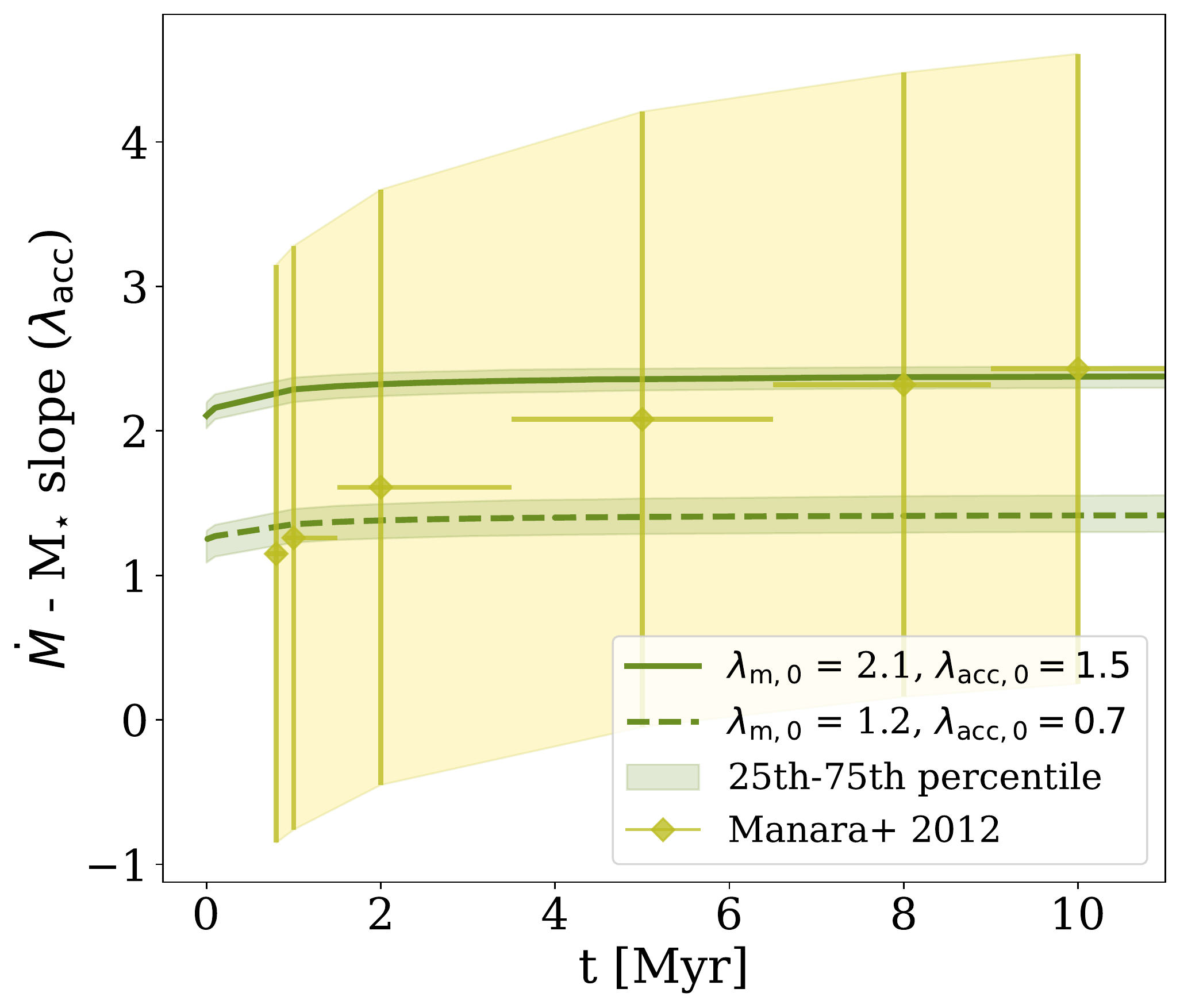}
    \caption{Same as Figure \ref{fig:data_comp_mdot_Testi}, but compared to the observational data from \protect\citetalias{Manara+2012}. See the caption of Figure \ref{fig:data_comp_mdisc} for detail on annotations. Note that the horizontal bars in this figure do not represent the errors on the ages, but rather the extent of the age interval that was considered in performing the fit (see text for details).}
    \label{fig:data_comp_mdot_Manara}
\end{figure}

We also analysed the time evolution in the spreads of both correlations. The spread is determined as the standard deviation of the vertical distances of every point from the fitted relation. Figure \ref{fig:sigmaevo} shows the results with input values $\sigma_{M}(0) = 0.65$ dex, $\sigma_{R} (0) = 0.52$ dex (chosen to be consistent with the observed values in the data sets from \citetalias{Ansdell+2017} and \citetalias{Testi+2022-Ofiucone}, referring to log-normal distributions). The light blue and pink line show the evolution of the dispersion of the $M_{\mathrm{d}} - M_{\star}$ and $\dot M - M_{\star}$ correlations respectively. As we pointed out in paragraph \ref{subsub:evospread}, we expect the spread of the two correlations to reach the same evolved value, represented in the figure by the black dashed line. The numerical limit of 0.71 dex is very close to the theoretical one of 0.7; the slight difference is most likely due to the fact that the numerical dispersion in $t_{\nu}$ also accounts for the dispersion in stellar masses, which is not considered in the analytical calculation.

\section{Discussion} \label{sec:discussion}

In Section \ref{sec:analytical_considerations}, we have performed a theoretical analysis of the viscous evolution of power-law initial conditions. Through it, we have determined three possible physical scenarios which differ by the relative values of the initial slopes $\lambda_{\mathrm{m}, 0}$ and $\lambda_{\mathrm{acc}, 0}$. We have then analysed the output of our population synthesis code \texttt{Diskpop} and compared the evolutionary behaviour with our theoretical expectations (Section \ref{sec:population_synthesis}). In this Section, we perform a preliminary comparison with some of the available observational data (section \ref{subsec:obsdata}) and discuss the implications of our findings (section \ref{subsec:evoimplications}).

\subsection{Preliminary comparison to observational data} \label{subsec:obsdata}

After we tested \texttt{Diskpop}, recovering the analytical prediction for the viscous evolution of $\lambda_{\mathrm{m}}$ and $\lambda_{\mathrm{acc}}$, we performed a preliminary comparison using three sets of observational data (\citetalias{Ansdell+2017}, \citetalias{Testi+2022-Ofiucone}, \citetalias{Manara+2012}). 

There are some relevant caveats to these comparisons. First of all, we did not include any observational effects, biases, or sensitivity limitations in our simulations. However, the goal of this comparison is not to fit any data set, nor to perfectly reproduce the observations: our aim is just to have a qualitative idea of the parameter space, and to see whether the general trend shown by observations can (at least in principle) be reproduced from the theoretical point of view.

Figure \ref{fig:data_comp_mdisc}, \ref{fig:data_comp_mdot_Testi} and \ref{fig:data_comp_mdot_Manara} show the data comparison for the $M_{\mathrm{d}} - M_{\star}$ (Figure \ref{fig:data_comp_mdisc}) and the $\dot M - M_{\star}$ (Figure \ref{fig:data_comp_mdot_Testi} and \ref{fig:data_comp_mdot_Manara}) correlation. The top panel of Figure \ref{fig:data_comp_mdisc} refers to data from \citetalias{Ansdell+2017}, the bottom panel of Figure \ref{fig:data_comp_mdisc} and Figure \ref{fig:data_comp_mdot_Testi} from  \citetalias{Testi+2022-Ofiucone}, and Figure \ref{fig:data_comp_mdot_Manara} from \citetalias{Manara+2012}. Each panel in these figures contains two sets of simulations, represented by the solid and dashed lines respectively, corresponding to two different sets of parameters ($\lambda_{\mathrm{m}, 0}$ and $\lambda_{\mathrm{acc}, 0}$, see legend), with a spread of $\sigma_M(0) = 0.65$ dex and $\sigma_R(0) = 0.52$ dex. The darker shaded areas around the solid and dashed lines represent the 25th and 75th percentiles for the fitted slopes, obtained from 100 different statistical realisations of the same simulations. The initial conditions on $\lambda_{\mathrm{m}}$ and $\lambda_{\mathrm{acc}}$ were chosen to qualitatively match the observed slopes at later and younger ages respectively; the other parameters in the simulations are fixed - in particular, $H/R\big|_{R= 1 \text{ au}} = 0.33$ for $M_{\star} = 1 M_{\odot}$, $\alpha = 10^{-3}$, and $\beta = - 0.5$. The mean observational values are represented by diamonds, and the vertical and horizontal lines show their uncertainties. As we already discussed in Section \ref{sec:summary_obs}, there is a key difference between the data by \citetalias{Manara+2012} and all of the other observations: \citetalias{Manara+2012} focused on a single star-forming region, dividing the protostars based on their isochronal age group, while data from both \citetalias{Ansdell+2017} and \citetalias{Testi+2022-Ofiucone} instead refer to different star-forming regions, each with its own mean age. For this reason, the horizontal bars in Figure \ref{fig:data_comp_mdot_Manara} do not refer to the errors in age, but rather to the extent of the age intervals considered when performing the fits. Because of these considerable differences from the observational point of view, we divided Figure \ref{fig:data_comp_mdot_Testi} and \ref{fig:data_comp_mdot_Manara}.

It is worth recalling that, out of the two diagnostics $M_{\mathrm{d}}$ and $\dot M$, the second is more readily translated in the observational space. On the other hand, disc mass measurements always refer to \textit{dust} content; given that our simulations only include the gaseous component, all of the observed masses shown here have been multiplied by the dust-to-gas ratio (set to the usual value of 100, see \citealt{Bohlin+1978-dusttogas}). If \texttt{Diskpop} included dust evolution, the numerical disc mass $M_{\mathrm{d}}$ would be affected - and so would the $M_{\mathrm{d}} - M_{\star}$ correlation and its time evolution; on the other hand, we do not expect a significant influence of the presence of dust on the accretion rate. At this stage, we also assumed spread-less initial conditions for the disc masses and radii. For these reasons, as well as not including any observational effects or biases, the comparison shown in Figures \ref{fig:data_comp_mdisc}, \ref{fig:data_comp_mdot_Testi} and \ref{fig:data_comp_mdot_Manara} is qualitative.

Nonetheless, our comparison does suggest some interesting ideas. \citetalias{Ansdell+2017} claimed to have found an increasing trend of $\lambda_{\mathrm{m}}$ with the age of the sample, which can also apply to \citetalias{Testi+2022-Ofiucone} - even if it is not as evident. Furthermore, in \citetalias{Testi+2022-Ofiucone} the mean value for $\lambda_{\mathrm{acc}}$ seems to be increasing as well: the visible exception of the last point (Upper Scorpius) must be treated carefully, as it represents a highly incomplete sample - hence the large error bars (see the original paper for more detail). Due to the method adopted to determine the age of the subgroups in the Orion Nebula Cluster, \citetalias{Manara+2012} found significant errors on the fitted slopes; despite not being statistically significant, the mean measured values are increasing, and these data show a general agreement with the steepening trend. The interesting message of Figure \ref{fig:data_comp_mdisc}, \ref{fig:data_comp_mdot_Testi} and \ref{fig:data_comp_mdot_Manara} is that the observed increasing trend can be reproduced from the theoretical point of view, as the simulations show. Despite including a spread of around 0.5 dex in both the disc mass and radius, the analytical expectations are well recovered: given the choices of parameters (in particular, $\delta_0 > 1/2$) our simulations show the steepening evolution of both $\lambda_{\mathrm{m}}$ and $\lambda_{\mathrm{acc}}$. The main result is that, if this is the case, such observed correlations would be naturally explained in the purely viscous framework by

\begin{enumerate}
    \item assuming power-law correlations as initial conditions;
    \item imposing a positive correlation between the viscous timescale and the stellar mass.
\end{enumerate} 

Figure \ref{fig:data_comp_mdisc}, \ref{fig:data_comp_mdot_Testi} and \ref{fig:data_comp_mdot_Manara} also show the effect of a spread in the initial conditions on the evolution of the slopes. At $t = 0$, the median slopes (solid and dashed lines for the two initial conditions) and the input values $\lambda_{\mathrm{m}, 0}$ and $\lambda_{\mathrm{acc}, 0}$ coincide with a precision of $\sim 10^{-3}$ and $\sim 10^{-2}$ respectively; moreover, the evolutionary path of both $\lambda_{\mathrm{m}}$ and $\lambda_{\mathrm{acc}}$ is not changed. Therefore, including a spread in the initial condition only influences the starting point of the evolution, and does not affect the fundamental results that we have discussed.

As we mentioned above, we chose the parameters in the two different runs of \texttt{Diskpop} to qualitatively match the youngest and most evolved values for both $\lambda_{\mathrm{m}, \mathrm{obs}}$ and $\lambda_{\mathrm{acc}, \mathrm{obs}}$: this identifies a region of the parameter space for $\lambda_{\mathrm{m}, 0}$ and $\lambda_{\mathrm{acc}, 0}$, which lead to an evolution that roughly matches the observations (within their $1 \sigma$ error bars). These two sets of parameters are given by $\lambda_{\mathrm{m}, 0} = 2.1$, $\lambda_{\mathrm{acc}, 0} = 1.5$ and $\lambda_{\mathrm{m}, 0} = 1.2$, $\lambda_{\mathrm{acc}, 0} = 0.7$; the interval in the parameter space is

\begin{equation}
    \lambda_{\mathrm{m}, 0} \in [1.2, 2.1] \quad \lambda_{\mathrm{acc}, 0} \in [0.7, 1.5].
    \label{eq:interval_initcond}
\end{equation}

\noindent Figure \ref{fig:data_comp_mdisc}, \ref{fig:data_comp_mdot_Testi} and \ref{fig:data_comp_mdot_Manara} unveil another feature worth pointing out. Different values of $\lambda_{\mathrm{m}, 0}$ and $\lambda_{\mathrm{acc}, 0}$ can change the starting point of the curves obtained, but cannot change their shape - i.e., the rate of steepening or flattening of the slopes is not influenced by the initial parameters in the purely viscous case. We will discuss this further in the next Section.

\subsection{Evolution of the slopes}\label{subsec:evoimplications}

As we discussed above, the key promising aspect worth pointing out is that there is indeed a theoretical scenario which is able to reproduce the observed steepening of the correlations. Furthermore, this case also corresponds to the most likely physical regime - which leads to a positive correlation between the disc radius and the stellar mass. If these analytical requirements are matched, our analysis finds that power-law initial conditions for the correlations $\lambda_{\mathrm{m}}$ and $\lambda_{\mathrm{acc}}$ can only be steepened by viscous evolution. This implies that the observed correlations could be easily explained in terms of initial conditions and viscous evolution, without invoking any additional physical process. 

On the other hand, we also found that changing the initial values $\lambda_{\mathrm{m}, 0}$ and $\lambda_{\mathrm{acc}, 0}$ raises or lowers the evolution curve, adding only a small modification to the rate at which the exponents of the correlation steepen. Indeed, even if the difference $\delta_0 = \lambda{_{\mathrm{m}, 0}} - \lambda_{\mathrm{acc}, 0}$ does impact the viscous timescales, with these choices of parameters they still remain of the order of 1 Myr; as the bulk of the evolution in our models happens on comparable timescales, the effect of these changes is minimal.

Initial conditions within the parameters space specified in Equation \eqref{eq:interval_initcond}, and with $\delta_0 > 1/2$, are able to qualitatively match the observed slopes within $1 \sigma$ (although note that, as the error bars are large, this is not strongly constraining on any of the three models). However, the youngest and oldest populations seem to be better described by different simulations. This can be explained in two different ways:

\begin{enumerate}
    \item initial conditions are not the same for every star-forming region. This would mean that different curves must be considered for different regions to match their evolution;
    \item viscous evolution alone is not enough to explain the whole extent of star-forming regions.
\end{enumerate}

\noindent The second possibility is very much reasonable; a number of physical effects are now thought to affect disc evolution, sooner or later in their lifetime. We can mention in particular MHD winds, which are expected to play a role alongside viscosity \citep{Tabone+2021a}, internal and external photoevaporation \citep{ClarkeGendrinSoto2001, AlexanderClarkePringle2004, Owen+2012}, or dust evolution \citep{Sellek+2020-dustyorigin}; these effects are complicated to include in the mathematical analysis, but may be implemented in numerical codes to test whether they do lead to any modifications to the steepening rates of $\lambda_{\mathrm{m}}$ and $\lambda_{\mathrm{acc}}$. Dust evolution in particular would likely affect the computed disc mass, while it should not have a considerable influence on the accretion rate. This expected behaviour is encouraging, as Figure \ref{fig:data_comp_mdisc} shows that the shape of the time evolution of $\lambda_{\mathrm{acc}}$ (right panel) qualitatively matches the observed data way better than that of $\lambda_{\mathrm{m}}$ (left and central panels). An ideal candidate to show the signature of more complex evolutionary patterns would be Upper Scorpius, which is notorious for being difficult to explain through classic viscous models \citep{Manara+2020-UpperSco, Trapman+2020-COradii} unless adding additional physics, such as dust evolution \citep{Sellek+2020-dustyorigin}.

\section{Conclusions} \label{sec:conclusions}

In this paper, we investigated the correlation between the stellar mass and disc properties, $M_{\mathrm{d}} - M_{\star}$ and $\dot M - M_{\star}$. Assuming power-law correlations as initial conditions ($M_{\mathrm{d}}(0) \propto {M_{\star}}^{\lambda_{\mathrm{m}, 0}}$, $\dot M(0) \propto {M_{\star}}^{\lambda_{\mathrm{acc}, 0}}$), as well as viscous evolution, we obtained analytical equations to describe how the exponents of these correlations should evolve with time. Our main findings are the following:

\begin{enumerate}
    \item In the viscous picture, the two correlations follow the same trend: they both either grow steeper or shallower with time. Given enough time, they tend to the same value, which is determined by the initial slopes or, in other terms, by the scaling of the viscous timescale with the stellar mass. If we further make the plausible assumption that $H/R \propto {M_{\star}}^{\beta}$, with $\beta = - 1/2$  (as supported by radiative transfer models), $\alpha$ does not depend on $M_{\star}$ and $\delta_0 = \lambda_{\mathrm{m}, 0} - \lambda_{\mathrm{acc}, 0} > 1/2$, both correlations steepen with time, as tentatively suggested by observations. This is a consequence of viscous time increasing with stellar mass. Moreover, the spread of the two correlations is also bound to evolve to reach the same value, determined by the initial spreads in disc masses and viscous timescales;
    \item motivated by this early success, we attempted a comparison of our predictions with the available data. To this end we built a numerical tool, \texttt{Diskpop}, to evolve synthetic disc populations and include effects that cannot be tackled in the analytical approach (such as a spread in the initial conditions and the effect of random sampling);
    \item with the two sets of initial conditions $\lambda_{\mathrm{m}, 0} = 1.2, \lambda_{\mathrm{acc}, 0} = 0.7$ and $\lambda_{\mathrm{m}, 0} = 2.1, \lambda_{\mathrm{acc}, 0} = 1.5$ we obtain two limit curves which match either the youngest or the oldest region in the samples (within their error bars). This means that the range $\lambda_{\mathrm{m}, 0} \in [1.2, 2.1], \Hquad \lambda_{\mathrm{acc}, 0} \in [0.7, 1.5]$ as exponents in the initial conditions allows to span the observed parameter space;
    \item changing the initial condition can either raise or lower the curve, without significantly modifying its shape: this implies that our model cannot exactly match both the youngest and the oldest populations. This behaviour can imply two consequences: either the initial condition is not the same for every star-forming region, or viscosity alone is not enough to explain the observed long-term evolution (as already discussed in previous works, e.g. \cite{Mulders+2017}, \cite{Manara+2020-UpperSco}, \cite{Testi+2022-Ofiucone}). Of course, these two possibilities can also be valid at the same time. In particular, we expect dust evolution and planet formation, followed by disc-planet interaction, to play a crucial role: affecting the disc mass $M_{\mathrm{d}}$, it will likely change the time evolution of $\lambda_{\mathrm{m}}$. However, it would probably not affect $\lambda_{\mathrm{acc}}$ - which intriguingly already resembles the observations better.
\end{enumerate}

In our future work we aim at further developing \texttt{Diskpop}, increasing its complexity adding relevant physical effects such as MHD winds, photoevaporation and dust evolution; this will allow us to test the influence of these phenomena on the evolution of populations, and to obtain numerical results more suitable for a comparison with observational data.

\section{Data availability}

All the datasets generated and analysed during this study are avail- able from the corresponding author on reasonable request.

\section*{Acknowledgements}

We thank an anonymous referee for their valuable comments that helped us improve the clarity of this paper. We thank Richard Alexander, Barbara Ercolano, Til Birnstiel and Kees Dullemond for the interesting discussions. This work was partly supported by the Italian Ministero dell'Istruzione, Universit\`a e Ricerca through the grant Progetti Premiali 2012 – iALMA (CUP C$52$I$13000140001$), by the Deutsche Forschungs-gemeinschaft (DFG, German Research Foundation) - Ref no. 325594231 FOR $2634$/$1$ TE $1024$/$1$-$1$, by the DFG cluster of excellence Origins (www.origins-cluster.de), and from the European Research Council (ERC) via the ERC Synergy Grant {\em ECOGAL} (grant 855130). GR acknowledges support from an STFC Ernest Rutherford Fellowship (grant number ST/T003855/1). M.T. has been supported by the UK Science and Technology research Council (STFC) via the consolidated grant ST/S000623/1. G.L., G.R, L.T., C.F.M., and C.T. have been supported by the European Union’s Horizon 2020 research and innovation programme under the Marie Sklodowska-Curie grant agreement No. 823823 (RISE DUSTBUSTERS project).



\bibliographystyle{mnras}
\bibliography{bibliografia} 

\begin{thebibliography}{}
\makeatletter
\relax
\def\mn@urlcharsother{\let\do\@makeother \do\$\do\&\do\#\do\^\do\_\do\%\do\~}
\def\mn@doi{\begingroup\mn@urlcharsother \@ifnextchar [ {\mn@doi@}
  {\mn@doi@[]}}
\def\mn@doi@[#1]#2{\def\@tempa{#1}\ifx\@tempa\@empty \href
  {http://dx.doi.org/#2} {doi:#2}\else \href {http://dx.doi.org/#2} {#1}\fi
  \endgroup}
\def\mn@eprint#1#2{\mn@eprint@#1:#2::\@nil}
\def\mn@eprint@arXiv#1{\href {http://arxiv.org/abs/#1} {{\tt arXiv:#1}}}
\def\mn@eprint@dblp#1{\href {http://dblp.uni-trier.de/rec/bibtex/#1.xml}
  {dblp:#1}}
\def\mn@eprint@#1:#2:#3:#4\@nil{\def\@tempa {#1}\def\@tempb {#2}\def\@tempc
  {#3}\ifx \@tempc \@empty \let \@tempc \@tempb \let \@tempb \@tempa \fi \ifx
  \@tempb \@empty \def\@tempb {arXiv}\fi \@ifundefined
  {mn@eprint@\@tempb}{\@tempb:\@tempc}{\expandafter \expandafter \csname
  mn@eprint@\@tempb\endcsname \expandafter{\@tempc}}}

\bibitem[\protect\citeauthoryear{{Alcal{\'a}} et~al.,}{{Alcal{\'a}}
  et~al.}{2014}]{Alcala+2014}
{Alcal{\'a}} J.~M.,  et~al., 2014, \mn@doi [\aap]
  {10.1051/0004-6361/201322254}, \href
  {https://ui.adsabs.harvard.edu/abs/2014A&A...561A...2A} {561, A2}

\bibitem[\protect\citeauthoryear{{Alcal{\'a}} et~al.,}{{Alcal{\'a}}
  et~al.}{2017}]{Alcala+2017}
{Alcal{\'a}} J.~M.,  et~al., 2017, \mn@doi [\aap]
  {10.1051/0004-6361/201629929}, \href
  {https://ui.adsabs.harvard.edu/abs/2017A&A...600A..20A} {600, A20}

\bibitem[\protect\citeauthoryear{{Alexander} \& {Armitage}}{{Alexander} \&
  {Armitage}}{2006}]{AlexanderArmitage2006}
{Alexander} R.~D.,  {Armitage} P.~J.,  2006, \mn@doi [\apjl] {10.1086/503030},
  \href {https://ui.adsabs.harvard.edu/abs/2006ApJ...639L..83A} {639, L83}

\bibitem[\protect\citeauthoryear{{Alexander}, {Clarke}  \&
  {Pringle}}{{Alexander} et~al.}{2004}]{AlexanderClarkePringle2004}
{Alexander} R.~D.,  {Clarke} C.~J.,   {Pringle} J.~E.,  2004, \mn@doi [\mnras]
  {10.1111/j.1365-2966.2004.08161.x}, \href
  {https://ui.adsabs.harvard.edu/abs/2004MNRAS.354...71A} {354, 71}

\bibitem[\protect\citeauthoryear{{Andrews}, {Terrell}, {Tripathi}, {Ansdell},
  {Williams}  \& {Wilner}}{{Andrews} et~al.}{2018}]{Andrews2018Sizes}
{Andrews} S.~M.,  {Terrell} M.,  {Tripathi} A.,  {Ansdell} M.,  {Williams}
  J.~P.,   {Wilner} D.~J.,  2018, \mn@doi [\apj] {10.3847/1538-4357/aadd9f},
  \href {https://ui.adsabs.harvard.edu/abs/2018ApJ...865..157A} {865, 157}

\bibitem[\protect\citeauthoryear{{Ansdell} et~al.,}{{Ansdell}
  et~al.}{2016}]{Ansdell+2016Survey}
{Ansdell} M.,  et~al., 2016, \mn@doi [\apj] {10.3847/0004-637X/828/1/46}, \href
  {https://ui.adsabs.harvard.edu/abs/2016ApJ...828...46A} {828, 46}

\bibitem[\protect\citeauthoryear{{Ansdell}, {Williams}, {Manara}, {Miotello},
  {Facchini}, {van der Marel}, {Testi}  \& {van Dishoeck}}{{Ansdell}
  et~al.}{2017}]{Ansdell+2017}
{Ansdell} M.,  {Williams} J.~P.,  {Manara} C.~F.,  {Miotello} A.,  {Facchini}
  S.,  {van der Marel} N.,  {Testi} L.,   {van Dishoeck} E.~F.,  2017, \mn@doi
  [\aj] {10.3847/1538-3881/aa69c0}, \href
  {https://ui.adsabs.harvard.edu/abs/2017AJ....153..240A} {153, 240}

\bibitem[\protect\citeauthoryear{{Ansdell} et~al.,}{{Ansdell}
  et~al.}{2018}]{Ansdell+2018-radiigas}
{Ansdell} M.,  et~al., 2018, \mn@doi [\apj] {10.3847/1538-4357/aab890}, \href
  {https://ui.adsabs.harvard.edu/abs/2018ApJ...859...21A} {859, 21}

\bibitem[\protect\citeauthoryear{{Bai}}{{Bai}}{2017}]{Bai2017simMHD}
{Bai} X.-N.,  2017, \mn@doi [\apj] {10.3847/1538-4357/aa7dda}, \href
  {https://ui.adsabs.harvard.edu/abs/2017ApJ...845...75B} {845, 75}

\bibitem[\protect\citeauthoryear{{Barenfeld}, {Carpenter}, {Ricci}  \&
  {Isella}}{{Barenfeld} et~al.}{2016}]{Barenfeld2016-masssurvey}
{Barenfeld} S.~A.,  {Carpenter} J.~M.,  {Ricci} L.,   {Isella} A.,  2016,
  \mn@doi [\apj] {10.3847/0004-637X/827/2/142}, \href
  {https://ui.adsabs.harvard.edu/abs/2016ApJ...827..142B} {827, 142}

\bibitem[\protect\citeauthoryear{{B{\'e}thune}, {Lesur}  \&
  {Ferreira}}{{B{\'e}thune} et~al.}{2017}]{Bethune+2017}
{B{\'e}thune} W.,  {Lesur} G.,   {Ferreira} J.,  2017, \mn@doi [\aap]
  {10.1051/0004-6361/201630056}, \href
  {https://ui.adsabs.harvard.edu/abs/2017A&A...600A..75B} {600, A75}

\bibitem[\protect\citeauthoryear{{Bohlin}, {Savage}  \& {Drake}}{{Bohlin}
  et~al.}{1978}]{Bohlin+1978-dusttogas}
{Bohlin} R.~C.,  {Savage} B.~D.,   {Drake} J.~F.,  1978, \mn@doi [\apj]
  {10.1086/156357}, \href
  {https://ui.adsabs.harvard.edu/abs/1978ApJ...224..132B} {224, 132}

\bibitem[\protect\citeauthoryear{{Booth}, {Clarke}, {Madhusudhan}  \&
  {Ilee}}{{Booth} et~al.}{2017}]{Booth+2017-viscouscode}
{Booth} R.~A.,  {Clarke} C.~J.,  {Madhusudhan} N.,   {Ilee} J.~D.,  2017,
  \mn@doi [\mnras] {10.1093/mnras/stx1103}, \href
  {https://ui.adsabs.harvard.edu/abs/2017MNRAS.469.3994B} {469, 3994}

\bibitem[\protect\citeauthoryear{{Calvet} \& {Gullbring}}{{Calvet} \&
  {Gullbring}}{1998}]{CalvetGullbring1998}
{Calvet} N.,  {Gullbring} E.,  1998, \mn@doi [\apj] {10.1086/306527}, \href
  {https://ui.adsabs.harvard.edu/abs/1998ApJ...509..802C} {509, 802}

\bibitem[\protect\citeauthoryear{{Cazzoletti} et~al.,}{{Cazzoletti}
  et~al.}{2019}]{Cazzoletti2019}
{Cazzoletti} P.,  et~al., 2019, \mn@doi [\aap] {10.1051/0004-6361/201935273},
  \href {https://ui.adsabs.harvard.edu/abs/2019A&A...626A..11C} {626, A11}

\bibitem[\protect\citeauthoryear{{Cieza} et~al.,}{{Cieza}
  et~al.}{2019}]{Cieza+2019-OphSurvey}
{Cieza} L.~A.,  et~al., 2019, \mn@doi [\mnras] {10.1093/mnras/sty2653}, \href
  {https://ui.adsabs.harvard.edu/abs/2019MNRAS.482..698C} {482, 698}

\bibitem[\protect\citeauthoryear{{Clarke} \& {Pringle}}{{Clarke} \&
  {Pringle}}{2006}]{Clarke+2006-MdotMstarinPMS}
{Clarke} C.~J.,  {Pringle} J.~E.,  2006, \mn@doi [\mnras]
  {10.1111/j.1745-3933.2006.00177.x}, \href
  {https://ui.adsabs.harvard.edu/abs/2006MNRAS.370L..10C} {370, L10}

\bibitem[\protect\citeauthoryear{{Clarke}, {Gendrin}  \& {Sotomayor}}{{Clarke}
  et~al.}{2001}]{ClarkeGendrinSoto2001}
{Clarke} C.~J.,  {Gendrin} A.,   {Sotomayor} M.,  2001, \mn@doi [\mnras]
  {10.1046/j.1365-8711.2001.04891.x}, \href
  {https://ui.adsabs.harvard.edu/abs/2001MNRAS.328..485C} {328, 485}

\bibitem[\protect\citeauthoryear{{Cox} et~al.,}{{Cox} et~al.}{2017}]{Cox+2017}
{Cox} E.~G.,  et~al., 2017, \mn@doi [\apj] {10.3847/1538-4357/aa97e2}, \href
  {https://ui.adsabs.harvard.edu/abs/2017ApJ...851...83C} {851, 83}

\bibitem[\protect\citeauthoryear{{Dullemond}, {Natta}  \& {Testi}}{{Dullemond}
  et~al.}{2006}]{DullemondNattaTesti2006}
{Dullemond} C.~P.,  {Natta} A.,   {Testi} L.,  2006, \mn@doi [\apjl]
  {10.1086/505744}, \href
  {https://ui.adsabs.harvard.edu/abs/2006ApJ...645L..69D} {645, L69}

\bibitem[\protect\citeauthoryear{{Ercolano}, {Mayr}, {Owen}, {Rosotti}  \&
  {Manara}}{{Ercolano} et~al.}{2014}]{Ercolano+2014-MdotMstar}
{Ercolano} B.,  {Mayr} D.,  {Owen} J.~E.,  {Rosotti} G.,   {Manara} C.~F.,
  2014, \mn@doi [\mnras] {10.1093/mnras/stt2405}, \href
  {https://ui.adsabs.harvard.edu/abs/2014MNRAS.439..256E} {439, 256}

\bibitem[\protect\citeauthoryear{{Gullbring}, {Hartmann}, {Brice{\~n}o}  \&
  {Calvet}}{{Gullbring} et~al.}{1998}]{Gullbring1998}
{Gullbring} E.,  {Hartmann} L.,  {Brice{\~n}o} C.,   {Calvet} N.,  1998,
  \mn@doi [\apj] {10.1086/305032}, \href
  {https://ui.adsabs.harvard.edu/abs/1998ApJ...492..323G} {492, 323}

\bibitem[\protect\citeauthoryear{{Hartmann}, {Calvet}, {Gullbring}  \&
  {D'Alessio}}{{Hartmann} et~al.}{1998}]{Hartmann1998}
{Hartmann} L.,  {Calvet} N.,  {Gullbring} E.,   {D'Alessio} P.,  1998, \mn@doi
  [\apj] {10.1086/305277}, \href
  {https://ui.adsabs.harvard.edu/abs/1998ApJ...495..385H} {495, 385}

\bibitem[\protect\citeauthoryear{{Hendler}, {Pascucci}, {Pinilla}, {Tazzari},
  {Carpenter}, {Malhotra}  \& {Testi}}{{Hendler} et~al.}{2020}]{Hendler2020}
{Hendler} N.,  {Pascucci} I.,  {Pinilla} P.,  {Tazzari} M.,  {Carpenter} J.,
  {Malhotra} R.,   {Testi} L.,  2020, \mn@doi [\apj]
  {10.3847/1538-4357/ab70ba}, \href
  {https://ui.adsabs.harvard.edu/abs/2020ApJ...895..126H} {895, 126}

\bibitem[\protect\citeauthoryear{{Herczeg} \& {Hillenbrand}}{{Herczeg} \&
  {Hillenbrand}}{2008}]{Herczeg&Hillenbrand2008}
{Herczeg} G.~J.,  {Hillenbrand} L.~A.,  2008, \mn@doi [\apj] {10.1086/586728},
  \href {https://ui.adsabs.harvard.edu/abs/2008ApJ...681..594H} {681, 594}

\bibitem[\protect\citeauthoryear{{Kalari} et~al.,}{{Kalari}
  et~al.}{2015}]{Kalari+2015}
{Kalari} V.~M.,  et~al., 2015, \mn@doi [\mnras] {10.1093/mnras/stv1676}, \href
  {https://ui.adsabs.harvard.edu/abs/2015MNRAS.453.1026K} {453, 1026}

\bibitem[\protect\citeauthoryear{{Kroupa}}{{Kroupa}}{2001}]{Kroupa2001}
{Kroupa} P.,  2001, \mn@doi [\mnras] {10.1046/j.1365-8711.2001.04022.x}, \href
  {https://ui.adsabs.harvard.edu/abs/2001MNRAS.322..231K} {322, 231}

\bibitem[\protect\citeauthoryear{{Kurtovic} et~al.,}{{Kurtovic}
  et~al.}{2021}]{Kurtovic+2021}
{Kurtovic} N.~T.,  et~al., 2021, \mn@doi [\aap] {10.1051/0004-6361/202038983},
  \href {https://ui.adsabs.harvard.edu/abs/2021A&A...645A.139K} {645, A139}

\bibitem[\protect\citeauthoryear{{Lesur}}{{Lesur}}{2020}]{Lesur2020reviewMHD}
{Lesur} G.,  2020, arXiv e-prints, \href
  {https://ui.adsabs.harvard.edu/abs/2020arXiv200715967L} {p. arXiv:2007.15967}

\bibitem[\protect\citeauthoryear{{Lesur}, {Kunz}  \& {Fromang}}{{Lesur}
  et~al.}{2014}]{Lesur+2014}
{Lesur} G.,  {Kunz} M.~W.,   {Fromang} S.,  2014, \mn@doi [\aap]
  {10.1051/0004-6361/201423660}, \href
  {https://ui.adsabs.harvard.edu/abs/2014A&A...566A..56L} {566, A56}

\bibitem[\protect\citeauthoryear{{Liu}, {Jin}, {Li}, {Isella}  \& {Li}}{{Liu}
  et~al.}{2018}]{Liu+2018AlphaVarying}
{Liu} S.-F.,  {Jin} S.,  {Li} S.,  {Isella} A.,   {Li} H.,  2018, \mn@doi
  [\apj] {10.3847/1538-4357/aab718}, \href
  {https://ui.adsabs.harvard.edu/abs/2018ApJ...857...87L} {857, 87}

\bibitem[\protect\citeauthoryear{{Lodato}, {Scardoni}, {Manara}  \&
  {Testi}}{{Lodato} et~al.}{2017}]{Lodato2017}
{Lodato} G.,  {Scardoni} C.~E.,  {Manara} C.~F.,   {Testi} L.,  2017, \mn@doi
  [\mnras] {10.1093/mnras/stx2273}, \href
  {https://ui.adsabs.harvard.edu/abs/2017MNRAS.472.4700L} {472, 4700}

\bibitem[\protect\citeauthoryear{{Long} et~al.,}{{Long}
  et~al.}{2019}]{Long+2019}
{Long} F.,  et~al., 2019, \mn@doi [\apj] {10.3847/1538-4357/ab2d2d}, \href
  {https://ui.adsabs.harvard.edu/abs/2019ApJ...882...49L} {882, 49}

\bibitem[\protect\citeauthoryear{{Lynden-Bell} \& {Pringle}}{{Lynden-Bell} \&
  {Pringle}}{1974}]{LyndenBellPringle1974}
{Lynden-Bell} D.,  {Pringle} J.~E.,  1974, \mn@doi [\mnras]
  {10.1093/mnras/168.3.603}, \href
  {https://ui.adsabs.harvard.edu/abs/1974MNRAS.168..603L} {168, 603}

\bibitem[\protect\citeauthoryear{{Mamajek}}{{Mamajek}}{2009}]{Mamajek2009}
{Mamajek} E.~E.,  2009, in {Usuda} T.,  {Tamura} M.,   {Ishii} M.,  eds,
  American Institute of Physics Conference Series Vol. 1158, Exoplanets and
  Disks: Their Formation and Diversity. pp 3--10 (\mn@eprint {arXiv}
  {0906.5011}), \mn@doi{10.1063/1.3215910}

\bibitem[\protect\citeauthoryear{{Manara}, {Robberto}, {Da Rio}, {Lodato},
  {Hillenbrand}, {Stassun}  \& {Soderblom}}{{Manara}
  et~al.}{2012}]{Manara+2012}
{Manara} C.~F.,  {Robberto} M.,  {Da Rio} N.,  {Lodato} G.,  {Hillenbrand}
  L.~A.,  {Stassun} K.~G.,   {Soderblom} D.~R.,  2012, \mn@doi [\apj]
  {10.1088/0004-637X/755/2/154}, \href
  {https://ui.adsabs.harvard.edu/abs/2012ApJ...755..154M} {755, 154}

\bibitem[\protect\citeauthoryear{{Manara}, {Fedele}, {Herczeg}  \&
  {Teixeira}}{{Manara} et~al.}{2016a}]{Manara+2016-ChaI}
{Manara} C.~F.,  {Fedele} D.,  {Herczeg} G.~J.,   {Teixeira} P.~S.,  2016a,
  \mn@doi [\aap] {10.1051/0004-6361/201527224}, \href
  {https://ui.adsabs.harvard.edu/abs/2016A&A...585A.136M} {585, A136}

\bibitem[\protect\citeauthoryear{{Manara} et~al.,}{{Manara}
  et~al.}{2016b}]{Manara2016-evidenceacc}
{Manara} C.~F.,  et~al., 2016b, \mn@doi [\aap] {10.1051/0004-6361/201628549},
  \href {https://ui.adsabs.harvard.edu/abs/2016A&A...591L...3M} {591, L3}

\bibitem[\protect\citeauthoryear{{Manara} et~al.,}{{Manara}
  et~al.}{2017}]{Manara2017-Cha}
{Manara} C.~F.,  et~al., 2017, \mn@doi [\aap] {10.1051/0004-6361/201630147},
  \href {https://ui.adsabs.harvard.edu/abs/2017A&A...604A.127M} {604, A127}

\bibitem[\protect\citeauthoryear{{Manara} et~al.,}{{Manara}
  et~al.}{2020}]{Manara+2020-UpperSco}
{Manara} C.~F.,  et~al., 2020, \mn@doi [\aap] {10.1051/0004-6361/202037949},
  \href {https://ui.adsabs.harvard.edu/abs/2020A&A...639A..58M} {639, A58}

\bibitem[\protect\citeauthoryear{{Mohanty}, {Jayawardhana}  \&
  {Basri}}{{Mohanty} et~al.}{2005}]{Mohanty+2005}
{Mohanty} S.,  {Jayawardhana} R.,   {Basri} G.,  2005, \mn@doi [\apj]
  {10.1086/429794}, \href
  {https://ui.adsabs.harvard.edu/abs/2005ApJ...626..498M} {626, 498}

\bibitem[\protect\citeauthoryear{{Morbidelli}, {Lunine}, {O'Brien}, {Raymond}
  \& {Walsh}}{{Morbidelli} et~al.}{2012}]{Morbidelli2012-planetformreview}
{Morbidelli} A.,  {Lunine} J.~I.,  {O'Brien} D.~P.,  {Raymond} S.~N.,   {Walsh}
  K.~J.,  2012, \mn@doi [Annual Review of Earth and Planetary Sciences]
  {10.1146/annurev-earth-042711-105319}, \href
  {https://ui.adsabs.harvard.edu/abs/2012AREPS..40..251M} {40, 251}

\bibitem[\protect\citeauthoryear{{Mordasini}, {Molli{\`e}re}, {Dittkrist},
  {Jin}  \& {Alibert}}{{Mordasini} et~al.}{2015}]{Mordasini+2015}
{Mordasini} C.,  {Molli{\`e}re} P.,  {Dittkrist} K.~M.,  {Jin} S.,   {Alibert}
  Y.,  2015, \mn@doi [International Journal of Astrobiology]
  {10.1017/S1473550414000263}, \href
  {https://ui.adsabs.harvard.edu/abs/2015IJAsB..14..201M} {14, 201}

\bibitem[\protect\citeauthoryear{{Mulders}, {Pascucci}, {Manara}, {Testi},
  {Herczeg}, {Henning}, {Mohanty}  \& {Lodato}}{{Mulders}
  et~al.}{2017}]{Mulders+2017}
{Mulders} G.~D.,  {Pascucci} I.,  {Manara} C.~F.,  {Testi} L.,  {Herczeg}
  G.~J.,  {Henning} T.,  {Mohanty} S.,   {Lodato} G.,  2017, \mn@doi [\apj]
  {10.3847/1538-4357/aa8906}, \href
  {https://ui.adsabs.harvard.edu/abs/2017ApJ...847...31M} {847, 31}

\bibitem[\protect\citeauthoryear{{Muzerolle}, {Hillenbrand}, {Brice{\~n}o},
  {Calvet}  \& {Hartmann}}{{Muzerolle} et~al.}{2003}]{Muzerolle+2003}
{Muzerolle} J.,  {Hillenbrand} L.,  {Brice{\~n}o} C.,  {Calvet} N.,
  {Hartmann} L.,  2003, in Brown Dwarfs. p.~141

\bibitem[\protect\citeauthoryear{{Natta}, {Testi}, {Muzerolle}, {Randich},
  {Comer{\'o}n}  \& {Persi}}{{Natta} et~al.}{2004}]{Natta+2004-lowmassaccrates}
{Natta} A.,  {Testi} L.,  {Muzerolle} J.,  {Randich} S.,  {Comer{\'o}n} F.,
  {Persi} P.,  2004, \mn@doi [\aap] {10.1051/0004-6361:20040356}, \href
  {https://ui.adsabs.harvard.edu/abs/2004A&A...424..603N} {424, 603}

\bibitem[\protect\citeauthoryear{{Natta}, {Testi}  \& {Randich}}{{Natta}
  et~al.}{2006}]{Natta+2006}
{Natta} A.,  {Testi} L.,   {Randich} S.,  2006, \mn@doi [\aap]
  {10.1051/0004-6361:20054706}, \href
  {https://ui.adsabs.harvard.edu/abs/2006A&A...452..245N} {452, 245}

\bibitem[\protect\citeauthoryear{{Owen}, {Clarke}  \& {Ercolano}}{{Owen}
  et~al.}{2012}]{Owen+2012}
{Owen} J.~E.,  {Clarke} C.~J.,   {Ercolano} B.,  2012, \mn@doi [\mnras]
  {10.1111/j.1365-2966.2011.20337.x}, \href
  {https://ui.adsabs.harvard.edu/abs/2012MNRAS.422.1880O} {422, 1880}

\bibitem[\protect\citeauthoryear{{Pascucci} et~al.,}{{Pascucci}
  et~al.}{2016}]{Pascucci+2016}
{Pascucci} I.,  et~al., 2016, \mn@doi [\apj] {10.3847/0004-637X/831/2/125},
  \href {https://ui.adsabs.harvard.edu/abs/2016ApJ...831..125P} {831, 125}

\bibitem[\protect\citeauthoryear{{Pinilla}, {Pascucci}  \& {Marino}}{{Pinilla}
  et~al.}{2020}]{Pinilla+2020-MdiscMstar}
{Pinilla} P.,  {Pascucci} I.,   {Marino} S.,  2020, \mn@doi [\aap]
  {10.1051/0004-6361/201937003}, \href
  {https://ui.adsabs.harvard.edu/abs/2020A&A...635A.105P} {635, A105}

\bibitem[\protect\citeauthoryear{{Preibisch}}{{Preibisch}}{2012}]{Preibisch2021ReviewAges}
{Preibisch} T.,  2012, \mn@doi [Research in Astronomy and Astrophysics]
  {10.1088/1674-4527/12/1/001}, \href
  {https://ui.adsabs.harvard.edu/abs/2012RAA....12....1P} {12, 1}

\bibitem[\protect\citeauthoryear{{Rafikov}}{{Rafikov}}{2017}]{Rafikov2017}
{Rafikov} R.~R.,  2017, \mn@doi [\apj] {10.3847/1538-4357/aa6249}, \href
  {https://ui.adsabs.harvard.edu/abs/2017ApJ...837..163R} {837, 163}

\bibitem[\protect\citeauthoryear{{Rigliaco}, {Natta}, {Randich}, {Testi}  \&
  {Biazzo}}{{Rigliaco} et~al.}{2011}]{Rigliaco+2011}
{Rigliaco} E.,  {Natta} A.,  {Randich} S.,  {Testi} L.,   {Biazzo} K.,  2011,
  \mn@doi [\aap] {10.1051/0004-6361/201015299}, \href
  {https://ui.adsabs.harvard.edu/abs/2011A&A...525A..47R} {525, A47}

\bibitem[\protect\citeauthoryear{{Rosotti}, {Clarke}, {Manara}  \&
  {Facchini}}{{Rosotti} et~al.}{2017}]{Rosotti+2017}
{Rosotti} G.~P.,  {Clarke} C.~J.,  {Manara} C.~F.,   {Facchini} S.,  2017,
  \mn@doi [\mnras] {10.1093/mnras/stx595}, \href
  {https://ui.adsabs.harvard.edu/abs/2017MNRAS.468.1631R} {468, 1631}

\bibitem[\protect\citeauthoryear{{Rosotti}, {Booth}, {Tazzari}, {Clarke},
  {Lodato}  \& {Testi}}{{Rosotti} et~al.}{2019a}]{Rosotti+Letter2019}
{Rosotti} G.~P.,  {Booth} R.~A.,  {Tazzari} M.,  {Clarke} C.,  {Lodato} G.,
  {Testi} L.,  2019a, \mn@doi [\mnras] {10.1093/mnrasl/slz064}, \href
  {https://ui.adsabs.harvard.edu/abs/2019MNRAS.486L..63R} {486, L63}

\bibitem[\protect\citeauthoryear{{Rosotti}, {Tazzari}, {Booth}, {Testi},
  {Lodato}  \& {Clarke}}{{Rosotti} et~al.}{2019b}]{Rosotti+2019}
{Rosotti} G.~P.,  {Tazzari} M.,  {Booth} R.~A.,  {Testi} L.,  {Lodato} G.,
  {Clarke} C.,  2019b, \mn@doi [\mnras] {10.1093/mnras/stz1190}, \href
  {https://ui.adsabs.harvard.edu/abs/2019MNRAS.486.4829R} {486, 4829}

\bibitem[\protect\citeauthoryear{{Sanchis} et~al.,}{{Sanchis}
  et~al.}{2020}]{Sanchis+2020Lupus}
{Sanchis} E.,  et~al., 2020, \mn@doi [\aap] {10.1051/0004-6361/201936913},
  \href {https://ui.adsabs.harvard.edu/abs/2020A&A...633A.114S} {633, A114}

\bibitem[\protect\citeauthoryear{{Sanchis} et~al.,}{{Sanchis}
  et~al.}{2021}]{Sanchis2021}
{Sanchis} E.,  et~al., 2021, \mn@doi [\aap] {10.1051/0004-6361/202039733},
  \href {https://ui.adsabs.harvard.edu/abs/2021A&A...649A..19S} {649, A19}

\bibitem[\protect\citeauthoryear{{Sellek}, {Booth}  \& {Clarke}}{{Sellek}
  et~al.}{2020a}]{Sellek+2020-dustyorigin}
{Sellek} A.~D.,  {Booth} R.~A.,   {Clarke} C.~J.,  2020a, \mn@doi [\mnras]
  {10.1093/mnras/stz3528}, \href
  {https://ui.adsabs.harvard.edu/abs/2020MNRAS.492.1279S} {492, 1279}

\bibitem[\protect\citeauthoryear{{Sellek}, {Booth}  \& {Clarke}}{{Sellek}
  et~al.}{2020b}]{Sellek+2020-extphoto}
{Sellek} A.~D.,  {Booth} R.~A.,   {Clarke} C.~J.,  2020b, \mn@doi [\mnras]
  {10.1093/mnras/staa2519}, \href
  {https://ui.adsabs.harvard.edu/abs/2020MNRAS.498.2845S} {498, 2845}

\bibitem[\protect\citeauthoryear{{Shakura} \& {Sunyaev}}{{Shakura} \&
  {Sunyaev}}{1973}]{ShakuraSunyaev1973}
{Shakura} N.~I.,  {Sunyaev} R.~A.,  1973, in {Bradt} H.,  {Giacconi} R.,  eds,
  IAU Symposium Vol. 55, X- and Gamma-Ray Astronomy. p.~155

\bibitem[\protect\citeauthoryear{{Sinclair}, {Rosotti}, {Juhasz}  \&
  {Clarke}}{{Sinclair} et~al.}{2020}]{Sinclair2020}
{Sinclair} C.~A.,  {Rosotti} G.~P.,  {Juhasz} A.,   {Clarke} C.~J.,  2020,
  \mn@doi [\mnras] {10.1093/mnras/staa539}, \href
  {https://ui.adsabs.harvard.edu/abs/2020MNRAS.493.3535S} {493, 3535}

\bibitem[\protect\citeauthoryear{{Soderblom}, {Hillenbrand}, {Jeffries},
  {Mamajek}  \& {Naylor}}{{Soderblom} et~al.}{2014}]{Soderblom+2014PPVI}
{Soderblom} D.~R.,  {Hillenbrand} L.~A.,  {Jeffries} R.~D.,  {Mamajek} E.~E.,
  {Naylor} T.,  2014, in {Beuther} H.,  {Klessen} R.~S.,  {Dullemond} C.~P.,
  {Henning} T.,  eds, Protostars and Planets VI. p.~219 (\mn@eprint {arXiv}
  {1311.7024}), \mn@doi{10.2458/azu\_uapress\_9780816531240-ch010}

\bibitem[\protect\citeauthoryear{{Somigliana}, {Toci}, {Lodato}, {Rosotti}  \&
  {Manara}}{{Somigliana} et~al.}{2020}]{Somigliana2020}
{Somigliana} A.,  {Toci} C.,  {Lodato} G.,  {Rosotti} G.,   {Manara} C.~F.,
  2020, \mn@doi [\mnras] {10.1093/mnras/stz3481}, \href
  {https://ui.adsabs.harvard.edu/abs/2020MNRAS.492.1120S} {492, 1120}

\bibitem[\protect\citeauthoryear{{Tabone}, {Rosotti}, {Cridland}, {Armitage}
  \& {Lodato}}{{Tabone} et~al.}{2021a}]{Tabone+2021a}
{Tabone} B.,  {Rosotti} G.~P.,  {Cridland} A.~J.,  {Armitage} P.~J.,   {Lodato}
  G.,  2021a, \mn@doi [\mnras] {10.1093/mnras/stab3442}, \href
  {https://ui.adsabs.harvard.edu/abs/2021MNRAS.tmp.3115T} {}

\bibitem[\protect\citeauthoryear{{Tabone}, {Rosotti}, {Lodato}, {Armitage},
  {Cridland}  \& {van Dishoeck}}{{Tabone} et~al.}{2021b}]{Tabone+2021b}
{Tabone} B.,  {Rosotti} G.~P.,  {Lodato} G.,  {Armitage} P.~J.,  {Cridland}
  A.~J.,   {van Dishoeck} E.~F.,  2021b, arXiv e-prints, \href
  {https://ui.adsabs.harvard.edu/abs/2021arXiv211114473T} {p. arXiv:2111.14473}

\bibitem[\protect\citeauthoryear{{Tazzari} et~al.,}{{Tazzari}
  et~al.}{2017}]{Tazzari+2017}
{Tazzari} M.,  et~al., 2017, \mn@doi [\aap] {10.1051/0004-6361/201730890},
  \href {https://ui.adsabs.harvard.edu/abs/2017A&A...606A..88T} {606, A88}

\bibitem[\protect\citeauthoryear{{Testi}, {Natta}, {Scholz}, {Tazzari}, {Ricci}
   \& {de Gregorio Monsalvo}}{{Testi} et~al.}{2016}]{Testi+2016}
{Testi} L.,  {Natta} A.,  {Scholz} A.,  {Tazzari} M.,  {Ricci} L.,   {de
  Gregorio Monsalvo} I.,  2016, \mn@doi [\aap] {10.1051/0004-6361/201628623},
  \href {https://ui.adsabs.harvard.edu/abs/2016A&A...593A.111T} {593, A111}

\bibitem[\protect\citeauthoryear{{Testi} et~al.,}{{Testi}
  et~al.}{2022}]{Testi+2022-Ofiucone}
{Testi} L.,  et~al., 2022, arXiv e-prints, \href
  {https://ui.adsabs.harvard.edu/abs/2022arXiv220104079T} {p. arXiv:2201.04079}

\bibitem[\protect\citeauthoryear{{Toci}, {Rosotti}, {Lodato}, {Testi}  \&
  {Trapman}}{{Toci} et~al.}{2021}]{Toci+2021}
{Toci} C.,  {Rosotti} G.,  {Lodato} G.,  {Testi} L.,   {Trapman} L.,  2021,
  \mn@doi [\mnras] {10.1093/mnras/stab2112}, \href
  {https://ui.adsabs.harvard.edu/abs/2021MNRAS.507..818T} {507, 818}

\bibitem[\protect\citeauthoryear{{Trapman}, {Facchini}, {Hogerheijde}, {van
  Dishoeck}  \& {Bruderer}}{{Trapman} et~al.}{2019}]{Trapman2019}
{Trapman} L.,  {Facchini} S.,  {Hogerheijde} M.~R.,  {van Dishoeck} E.~F.,
  {Bruderer} S.,  2019, \mn@doi [\aap] {10.1051/0004-6361/201834723}, \href
  {https://ui.adsabs.harvard.edu/abs/2019A&A...629A..79T} {629, A79}

\bibitem[\protect\citeauthoryear{{Trapman}, {Rosotti}, {Bosman}, {Hogerheijde}
  \& {van Dishoeck}}{{Trapman} et~al.}{2020}]{Trapman+2020-COradii}
{Trapman} L.,  {Rosotti} G.,  {Bosman} A.~D.,  {Hogerheijde} M.~R.,   {van
  Dishoeck} E.~F.,  2020, \mn@doi [\aap] {10.1051/0004-6361/202037673}, \href
  {https://ui.adsabs.harvard.edu/abs/2020A&A...640A...5T} {640, A5}

\bibitem[\protect\citeauthoryear{{Trapman}, {Tabone}, {Rosotti}  \&
  {Zhang}}{{Trapman} et~al.}{2021}]{Trapman+2021-MHDradii}
{Trapman} L.,  {Tabone} B.,  {Rosotti} G.,   {Zhang} K.,  2021, arXiv e-prints,
  \href {https://ui.adsabs.harvard.edu/abs/2021arXiv211200645T} {p.
  arXiv:2112.00645}

\bibitem[\protect\citeauthoryear{{Weidenschilling}}{{Weidenschilling}}{1977}]{Weidenschilling1977}
{Weidenschilling} S.~J.,  1977, \mn@doi [\mnras] {10.1093/mnras/180.2.57},
  \href {https://ui.adsabs.harvard.edu/abs/1977MNRAS.180...57W} {180, 57}

\bibitem[\protect\citeauthoryear{{Williams}, {Cieza}, {Hales}, {Ansdell},
  {Ruiz-Rodriguez}, {Casassus}, {Perez}  \& {Zurlo}}{{Williams}
  et~al.}{2019}]{Williams+2019-OphSurvey}
{Williams} J.~P.,  {Cieza} L.,  {Hales} A.,  {Ansdell} M.,  {Ruiz-Rodriguez}
  D.,  {Casassus} S.,  {Perez} S.,   {Zurlo} A.,  2019, \mn@doi [\apjl]
  {10.3847/2041-8213/ab1338}, \href
  {https://ui.adsabs.harvard.edu/abs/2019ApJ...875L...9W} {875, L9}

\makeatother
\end{thebibliography}




\appendix

\section{\texorpdfstring{$R_d - M_{\star}$}{rdmstar} correlation}\label{appendix-rdmstar}

\begin{figure}
    \centering
    \includegraphics[width = 7cm, keepaspectratio]{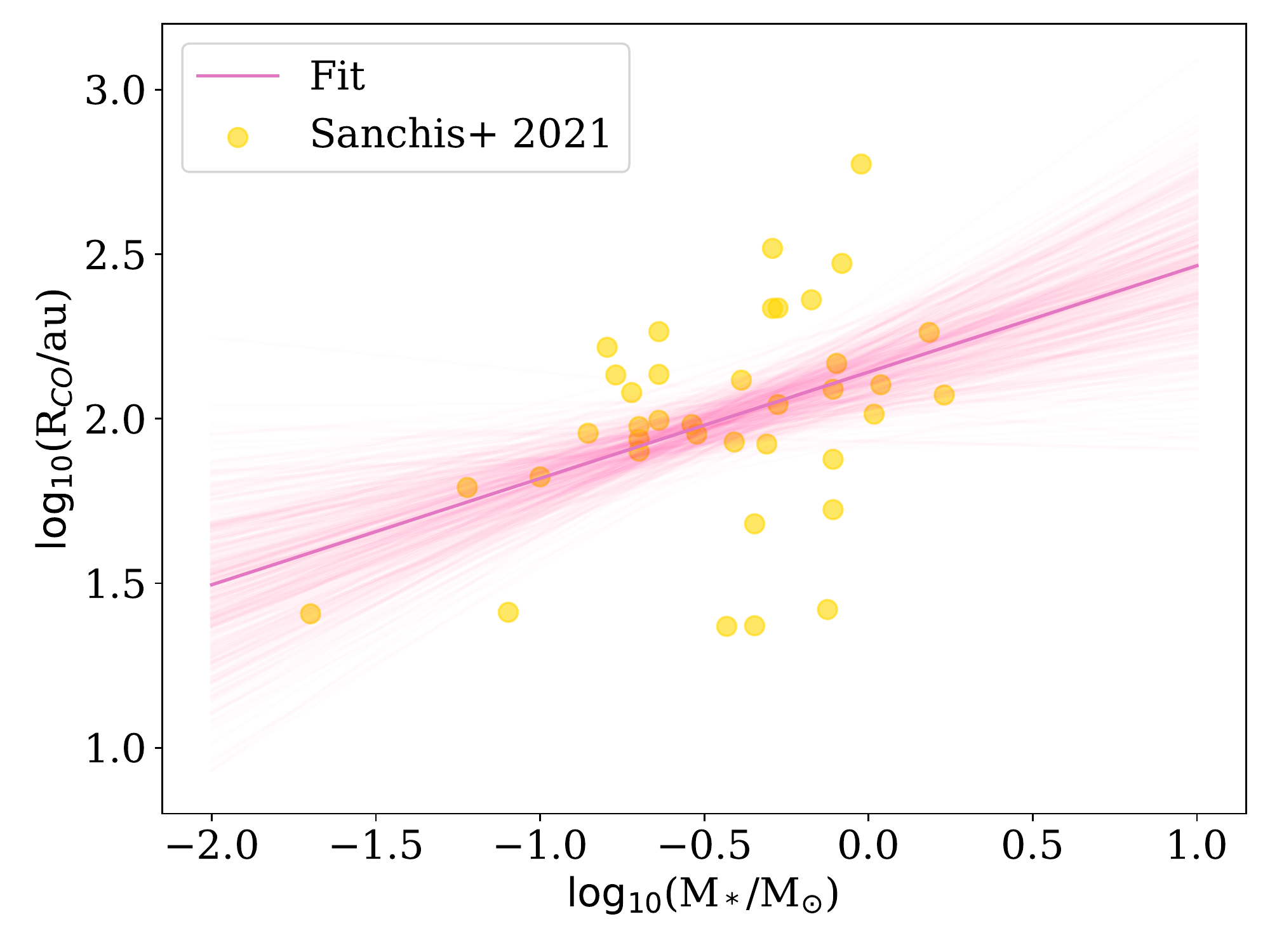}
    \caption{Radius obtained from CO observations (data by \protect\cite{Sanchis2021}) as a function of the stellar mass in the Lupus star-forming region.}
    \label{fig:sanchis}
\end{figure}

Figure \ref{fig:sanchis} shows a plot of the gas (CO) disc radius as a function of the stellar mass using data by \cite{Sanchis2021}, fitted using \texttt{linmix}. The best-fit parameters\footnote{Using the standard convention for linear relations, $y = \alpha + \beta x$.} are $\alpha = 2.14 \pm 0.08$, $\beta = 0.32 \pm 0.13$, with a spread $\sigma = 0.32 \pm 0.16$ dex. As discussed in paragraph \ref{subsec:disc_radius}, the correlation is positive but the fit is not particularly strong; the correlation coefficient is $\rho_{XY} = 0.32$, which implies weak correlation.

\section{Evolution of the disc mass distribution in the self-similar scenario}\label{appendix-spreads}

Here, we demonstrate that an initially log-normal distribution of disc masses keeps being log-normal at long times under the assumption of self-similar viscous evolution and estimate the relevant distribution parameters. 

We start by defining some useful quantities: let $m_0=\log M_{\rm d,0}$ and $\tau_\nu=\log t_\nu$ be the log of the initial disc mass and viscous time of a specific disc in a population. Now, the initial distributions of disc masses and viscous times are assumed to be log-normal, so that:

\begin{equation}
    \frac{\partial N}{\partial m_0}=N_1\exp[-(m_0-\bar{m}_0)^2/2\sigma_M^2],
\end{equation}

\begin{equation}
    \frac{\partial N}{\partial \tau_\nu}=N_2\exp[-(\tau_\nu-\bar{\tau}_\nu)^2/2\sigma_{t_\nu}^2],
\end{equation}

\noindent where $\bar{m}_0$ and $\bar{\tau}_\nu$ are the average values of the initial disc mass and viscous time (in logarithmic sense), $\sigma_M$ and $\sigma_{t_\nu}$ are their dispersions, and $N_1$ and $N_2$ are normalization constants. Note that the two log-normal probability distributions are independent, in the $m_0 - \tau_{\nu}$ space, and their means and variances may depend on the stellar mass.
For a self-similar disc in the limit $t\gg t_\nu$, the disc mass at time $t$ is $M_{\rm d}(t)=M_{\rm d,0}(t/t_\nu)^a$, where $a=1-\eta$ (see Eq. \ref{eq:discmass_ss}). In logarithmic form, this is:

\begin{equation}
    m=m_0+a\tau-a\tau_\nu, 
    \label{eq:appendix1}
\end{equation}

\noindent where $m=\log M_{\rm d}(t)$ and $\tau=\log t$. We can invert Eq. (\ref{eq:appendix1}), obtaining:

\begin{equation}
    \tau_{\nu,m}=\frac{m_0}{a}-\frac{m}{a}+\tau,
    \label{eq:appendix2}
\end{equation}

\noindent so that $\tau_{\nu,m}$ is the viscous time for which a disc with initial mass $m_0$ evolves into $m$ after a time $\tau$. 

The distribution of disc masses at time $t$ can then be obtained by integrating over the distribution of initial disc masses and viscous times, under the condition that Eq. (\ref{eq:appendix2}) is satisfied:

\begin{equation}
    \frac{\partial N}{\partial m}=\int\int \frac{\partial N}{\partial m_0}\frac{\partial N}{\partial \tau_\nu}\delta(\tau_\nu-\tau_{\nu,m})\mbox{d}\tau_\nu \mbox{d} m_0,
\end{equation}

\noindent where here $\delta$ is the Dirac function. The integral over viscous times can be readily be done, and, after inserting the initial distributions of disc masses and viscous times and using Eq. (\ref{eq:appendix2}), we obtain:

\begin{equation}
    \frac{\partial N}{\partial m} = N_1 N_2 \int \exp\left\{-\frac{(m_0-\bar m_0)^2}{2\sigma_M^2}-\left[ \frac{(m_0-m)}{a}+(\tau-\bar\tau_\nu)\right]^2\frac{1}{2\sigma_{t_\nu}^2}\right\}\mbox{d} m_0.
    \label{eq:appendix3}
\end{equation}

\noindent Let us define $\bar{m}=\bar{m}_0+a\tau-a\bar{\tau}_\nu$, which we will see is the average disc mass at time $\tau$. We also define $\tilde{m}=m-\bar{m}$ and $\tilde{m}_0=m_0-\bar{m}_0$. The exponent in Eq. \ref{eq:appendix3} can be then rewritten as (neglecting the factor $-1/2$):

\begin{equation}
    \mathcal{M}=\frac{\tilde{m}_0^2}{\sigma_M^2}+\left[\frac{\tilde{m}_0}{a}-\frac{\tilde{m}}{a}\right]^2\frac{1}{\sigma_{t_\nu}^2}.
\end{equation}

\noindent We can now easily rewrite the exponent $\mathcal{M}$ in such a way to isolate the dependence on $m_0$ (that we integrate upon). After some straightforward algebra, we get:

\begin{equation}
    \mathcal{M}=\left(A\tilde{m}_0-\frac{B^2}{A}\tilde{m}\right)^2 + \frac{A^2-B^2}{A^2/B^2}\tilde{m}^2,
    \label{eq:appendix4}
\end{equation}

\noindent where

\begin{equation}
    B^2=\frac{1}{a^2\sigma^2_{t_\nu}}
\end{equation}

\noindent and

\begin{equation}
    A^2=B^2+\frac{1}{\sigma^2_M}.
\end{equation}

\noindent Now, integrating Eq. (\ref{eq:appendix3}) over $\tilde{m}_0$, the whole first term on the RHS in Eq. (\ref{eq:appendix4}) translates into a normalization constant, leaving us with just a Gaussian distribution for $\tilde{m}$. This means that the disc masses at time $t$ are distributed log-normally, with mean $\bar{m}=\bar{m}_0+a\tau-a\bar{\tau}_\nu$ and with dispersion

\begin{equation}
    \sigma_M(t)=\frac{A^2/B^2}{A^2-B^2}=\sigma_M^2+a^2\sigma_{t_\nu}^2.
\end{equation}


\bsp	
\label{lastpage}
\end{document}